# Ultrafast Isomerization in Acetylene Dication After Carbon K-shell Ionization

Zheng Li,[1,2,*] Ludger Inhester,[3,4] Chelsea Liekhus-Schmaltz,[1,5] Basile F. E. Curchod,[1,2] James William Snyder, Jr.,[1,2] Nikita Medvedev,[3,8,9] James Cryan,[1] Timur Osipov,[1] Stefan Pabst,[6] Oriol Vendrell,[7] Phil Bucksbaum,[1,5] and Todd J. Martinez[1,2,‡]

[1]*SLAC National Accelerator Laboratory, 2575 Sand Hill Road, Menlo Park, CA 94025, USA*
[2]*Department of Chemistry and the PULSE Institute, Stanford University, 333 Campus Drive, Stanford, CA 94305, USA*
[3]*Center for Free Electron Laser Science, Deutsches Elektronen-Synchrotron, Notkestraße 85, D-22607 Hamburg, Germany*
[4]*Hamburg Center for Ultrafast Imaging, Luruper Chaussee 149, D-22761 Hamburg, Germany*
[5]*Department of Physics, Stanford University, 382 Via Pueblo Mall, Stanford, CA 94305, USA*
[6]*Harvard-Smithsonian Center for Astrophysics, 60 Garden Street, Cambridge, MA 02138, USA*
[7]*Department of Physics and Astronomy, Aarhus University, Ny Munkegade 120, 8000 Aarhus, Denmark*
[8]*Department of Radiation and Chemical Physics, Institute of Physics, Czech Academy of Sciences, Na Slovance 2, 182 21 Prague 8, Czech Republic*
[9]*Laser Plasma Department, Institute of Plasma Physics, Czech Academy of Sciences, Za Slovankou 3, 182 00 Prague 8, Czech Republic*

**Abstract:** Ultrafast proton migration and isomerization are key processes for acetylene and its ions. However the mechanism for ultrafast isomerization of acetylene [HCCH]$^{2+}$ to vinylidene [H$_2$CC]$^{2+}$ dication remains nebulous. Theoretical studies show a large potential barrier (>2eV) for isomerization on low-lying dicationic states, implying picosecond or longer isomerization timescales. However a recent experiment at a femtosecond X-ray free electron laser (XFEL) suggests sub-100fs isomerization. We address this contradiction with a complete theoretical study of the dynamics of acetylene dication produced by Auger decay after X-ray photoionization of the carbon atom *K* shell. We find no sub-100fs isomerization, while reproducing the salient features of the time-resolved Coulomb imaging experiment. This work resolves the seeming contradiction between experiment and theory and also calls for careful interpretation of structural information from the widely applied Coulomb momentum imaging method.

---

[*] zheng.li@cfel.de
[‡] toddjmartinez@stanford.edu

**Introduction**

Acetylene ($C_2H_2$) in neutral and ionic forms is an important species in combustion and atmospheric chemistry, and in the interstellar medium. The vinylidene isomer is an important intermediate in many reactions involving $C_2H_2$.[1-8] However, unlike in the neutral and cationic species,[6-11] the pathway for isomerization of the acetylene dication, consisting of hydrogen migration from $[HCCH]^{2+}$ to $[H_2CC]^{2+}$, remains largely unresolved. The reason for this disparity is an apparent contradiction between theory and experiment, prompting numerous studies.[5,12-18] Arguments on both sides can be summed up as follows. Experimental synchrotron data concludes that isomerization occurs on the low lying dicationic states $^1\Sigma_g$ and $^1\Delta_g$ with vacancies $1\pi_u^{-2}$, while deprotonation and symmetric breakup occur on the higher lying $^1\Pi_u$ states with $1\pi_u^{-1}3\sigma_g^{-1}$ character.[13,19] In addition, photoelectron-photoion momentum spectroscopy experiments[12,13] suggest that ultrafast hydrogen migration occurs in less than 100fs. A third separate XFEL experiment at Linac Coherent Light Source (LCLS) was interpreted to show the existence of significant hydrogen migration within 100fs.[5] These three pieces of evidence would seem to indicate that isomerization proceeds on the low lying dicationic states on a sub-100fs time scale.

In contrast to this experimental evidence, *ab initio* electronic structure calculations predict an isomerization barrier of ~2eV on the $1\pi_u^{-2}$ double hole states.[14,15] Isomerization over such a large barrier would be highly unlikely to occur on the femtosecond timescale and is expected to be orders of magnitude slower than the observed 100fs isomerization time. One possible solution to this conundrum is that while true isomerization occurs on the low lying dication states much more slowly than the experimental time scale, significant large-amplitude proton motion can occur on the sub-100fs time scale without leading to isomerization. As we will demonstrate, those dynamics can be easily misinterpreted as actual isomerization in Coulomb explosion measurements.

In the higher lying $1\pi_u^{-1}3\sigma_g^{-1}$ states of the dication, the isomerization channel is barrierless (see Supplementary Figure 4). However the double valence hole in these states weakens the C-C bond. The C-C bond of the dication therefore lengthens from 1.21Å to 1.37Å, with a relaxation energy of ~2.4 eV. The weak C-C bond coupled with the large energy release implies facile dissociation and fragmentation competes with isomerization. Even if there is no isomerization, significant large amplitude motion of the hydrogen atoms would be expected.

During or after the fragmentation, the fragments might rotate relative to each other. Upon Coulomb explosion, this relative rotation could masquerade as isomerization. As we will see, the experimental deuteron migration signal is well reproduced by our simulations. However, the signatures previously thought to arise from isomerization instead arise from the relative rotation of the fragments after breakup.

While the fragmentation and subsequent relative rotation of the fragments account for the major signal seen in the experiment, we did identify a sub-100 fs isomerization pathway that starts from a highly dissociative $1\pi_u^{-1}3\sigma_g^{-1}$ state and ends up with isomerized $[CH_2C]^{2+}$ on a $1\pi_u^{-2}$ state. This channel is predicted to be highly improbable, being observed only once out of 500 initial conditions and even then with a transition probability of less than $5 \times 10^{-4}$ (implying an estimated probability of less than $1 \times 10^{-6}$). Nevertheless, it is an interesting channel because it isomerizes through nonadiabatic transitions. Essentially, the molecule isomerizes on the $1\pi_u^{-1}3\sigma_g^{-1}$ state where torsion is facile. A nonadiabatic transition then takes the molecule to the lower lying $1\pi_u^{-2}$ state, on the vinylidene side of the isomerization barrier. This mechanism is only possible because of the breakdown of the Born-Oppenheimer approximation.

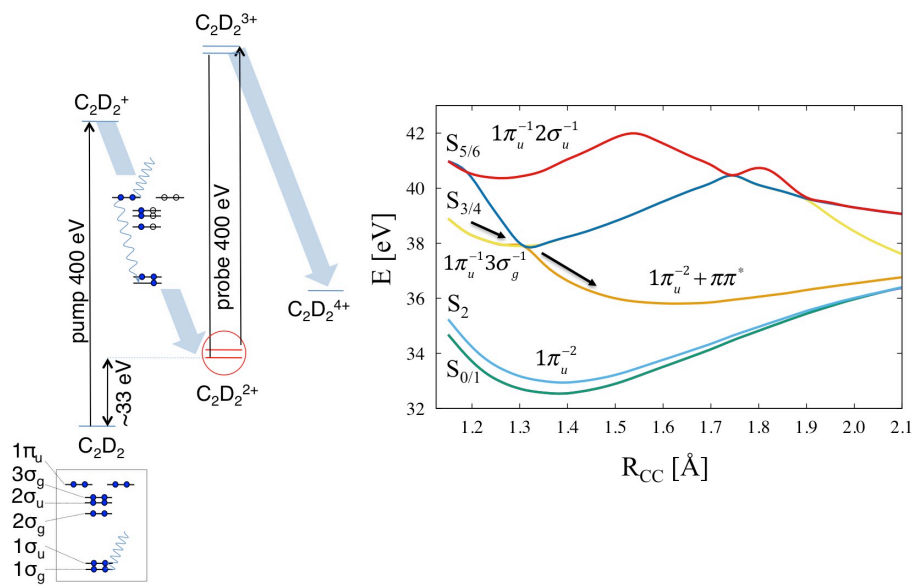

**Figure 1. Illustration of relevant dynamical processes after core ionization of acetylene.** Left: Sketch of the acetylene dication dynamics induced by X-ray photoionization and Auger decay. A first X-ray pump pulse core ionizes the neutral molecule to create the cation, which then undergoes Auger decay. A second X-ray probe pulse with a variable delay further core ionizes the dication, which promptly undergoes further Auger decay and Coulomb explosion. The momentum of the resulting fragments is measured to create the momentum map described in the text. Right: The potential curves of the singlet dicationic states are plotted in the adiabatic representation. The first 3 adiabatic states $S_0$-$S_2$ are dominated by a double hole configuration $1\pi_u^{-2}$, and the higher lying $S_3$ and $S_4$ states have the double hole configuration $1\pi_u^{-1}3\sigma_g^{-1}$, with one electron hole in each of the π- and σ-orbitals. The black arrows label the barrierless fragmentation pathway on $S_{3/4}$, arising from a crossing of the diabatic states $^1\Pi_u(1\pi_u^{-1}3\sigma_g^{-1})$ and $^1\Sigma_u(1\pi_u^{-2}+\pi_u\text{-}\pi_g^*)$.

In this work, we present a complete theoretical time-resolved picture of the ultrafast X-ray pump/X-ray probe experiment on acetylene dication dynamics. We model the dynamics of the core-ionized cation, its Auger decay, the dynamics of the dication, and the momentum distribution in the time-resolved Coulomb explosion imaging that was used to record a molecular movie for acetylene dication dynamics,[5] as shown schematically in Figure 1. Our results show that a sub 100fs isomerization in the low lying electronic states of the acetylene dication is unlikely.

**Results**

*Dynamics of the core ionized acetylene cation*

In our simulation, we follow the experiment shown schematically in Figure 1, and use $C_2D_2^+$ rather than $C_2H_2^+$. Deuterated acetylene was employed in the experiment to eliminate potential background sources of protons from water and other contaminants.[5] Although deuterated acetylene should have the same electronic structure except for negligible second order Born-Huang coupling, the particle velocity and the rate of tunneling through the barrier are expected to be slower in $C_2D_2^+$ compared with $C_2H_2^+$. The experiment was interpreted to show strong signatures of a vinylidene-like channel already in the first 12 fs following core ionization from Coulomb explosion momentum mapping,[5] implying that the X-ray induced dynamics starts immediately after photoionization of the carbon *K* shell. We model the dynamics on the core-hole state prior to the Auger decay that yields the dication. The potential energy surface of the core ionized cation on the $^2\Sigma_{g/u}$ state is calculated (during the dynamics) using a ΔSCF scheme with the maximum overlap method (details in Supplementary Note 1).[20] For modeling the core hole state, we adopt a localized picture with the core hole localized on one of the carbon atoms (i.e. breaking the gerade/ungerade symmetry). As shown in Figure 2a, the dynamical screening of the core hole through shake-up processes enhances the valence electron density in the C-C bond region. The electron reorganization strengthens the bonding along the molecular axis, and hardens the angular potential for the deuterons along the $\theta_{CCD}$ coordinate (as discussed in Supplementary Note 1). Our *ab initio* molecular dynamics simulations of core ionized acetylene cation start from initial conditions (positions and momenta of the atoms) sampled from the vibrational ground state harmonic Wigner distribution of neutral acetylene, as calculated with second-order perturbation theory (MP2) in the 6-31G* basis set.

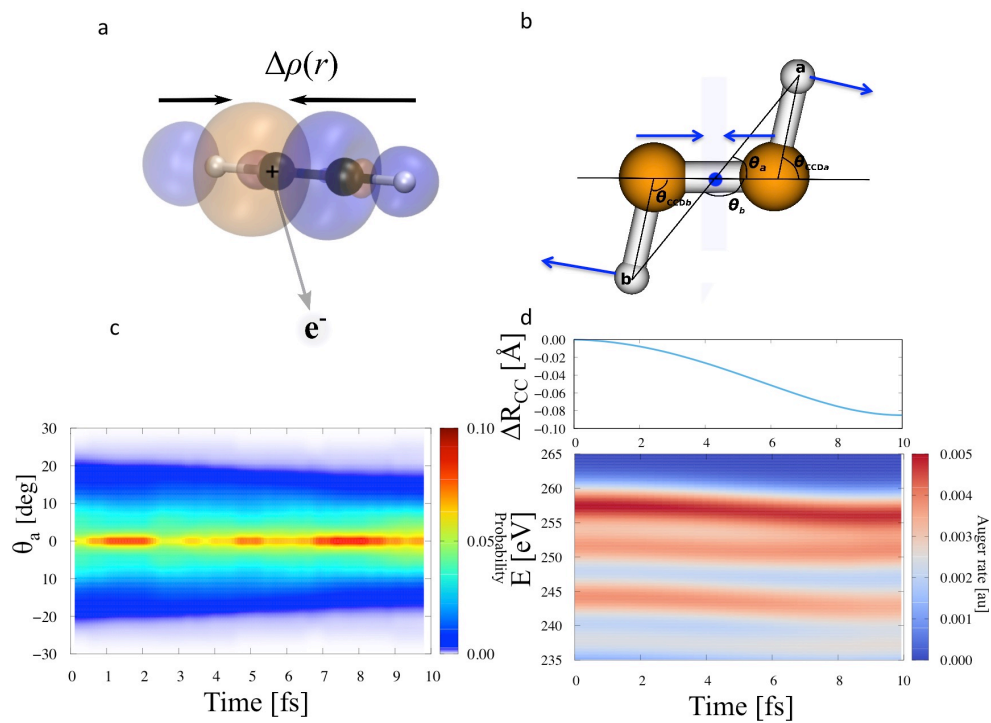

**Figure 2. Dynamics of the core ionized acetylene cation.** (a) The relaxed electron density $\rho_{-1}(r)$ after carbon $K$ edge photoionization. $\Delta\rho(r) = \rho_{-1}(r) - \tilde{\rho}_{-1}(r)$ is shown, where $\tilde{\rho}_{-1}(r)$ is the electron density of the unrelaxed core hole state after removing one electron from C$1s$ orbital. (b) Characteristic motion of cation, C-C bond contraction and CCD linearization. (c) Evolution of the CCD angle $\theta_a$ shows the cation evolves towards a narrower angular distribution. (d) The time-resolved Auger spectra from a representative trajectory and the evolution of C-C bond length and kinetic energy of Auger electrons. The contracting C-C bond results in a red shift of Auger electron energies. The Auger spectra are broadened by a Lorentzian corresponding to a core-hole lifetime of 8fs.

Figs. 2c and 2d present the evolution of the angular distribution of the deuterons and the C-C bond distance ($R_{C-C}$) after core-hole ionization by the X-ray pump pulse (on the cationic $^2\Sigma$ state). The nuclear dynamics after core ionization is rather limited and mainly characterized by a decrease in $R_{C-C}$ and narrowing of the CCD bending angle distribution. Along the core-hole ionized trajectories, we calculate the instantaneous Auger spectra (see Figure 2d and Supplmentary Note 2).[21] The energy of the emitted Auger electron is lowered with time because $R_{C-C}$ decreases on the cationic $^2\Sigma$ state (increasing the Coulomb repulsion of the two holes in the dication that results from Auger decay). Thus the energy difference between the core ionized state and the repulsive dicationic final state decreases as $R_{C-C}$ decreases. This effect leads to a redshifted Auger spectrum as the dynamics proceeds. Given the time-resolved Auger spectrum along each of the cationic trajectories, we populate the first five singlet and the first four triplet dicationic states with probabilities derived from the decay rate of the $K$ shell core hole state $|K\rangle$ to different dicationic states $|LL\rangle$ with double holes in the $L$-shell as $P_{KLL_i} = \rho_K(t)\Gamma_{KLL_i}(t)\Delta t$, where $\Gamma_{KLL_i}(t)$ is the Auger rate at a delay time of $t$ after C$1s$ core ionization and $\rho_K(t)$ is the population of the core ionized state determined by the kinetic rate equation $\dot{\rho}_K(t) = -\rho_K(t)\Gamma(t)$, $\rho_K(0) = 1$, where $\Gamma(t) = \sum_{|KLL_i\rangle} \Gamma_{KLL_i}(t)$ is the total Auger decay rate. We assume that the initial positions and momenta of the dication are inherited from the parent cation at the time of Auger decay and a recoil momentum from Auger electron is added on the carbon atom subject to primary core-ionization. In the following discussion, we focus on the lowest five singlet and four triplet dicationic states. Higher-lying states are not relevant because they are directly dissociative. Thus, population of these states will not lead to isomerization, but rather immediate fragmentation which is too fast to allow deuteron migration across the C-C bond. These higher lying states will therefore result in either symmetric breakup (CH$^+$+CH$^+$) or deprotonation (C$_2$H$^+$+H$^+$) channels. Additionally, the Auger decay rates into triplet states that support deuteron migration are one order of magnitude lower than those for the dominant singlet states. Thus, it is primarily the lowest five singlet states that are important to the question of isomerization in the X-ray pump experiments we discussed in the introduction[5,12,13] and we focus on these from here on. Generally the acetylene cation decays into C$_2$H$_2^{2+}$ dication with an Auger lifetime of ~8 fs.

*Acetylene Dication Dynamics*

We now focus on the dication dynamics initiated after Auger decay, and explore the possibility of isomerization within 100fs. Molecular dynamics on the dicationic states are described with the *ab initio* multiple spawning (AIMS) method.[22] At each time step, the electronic structure is solved using a state-averaged complete active space self-consistent field (CASSCF) wavefunction[23] with eight active electrons in eight orbitals using the 6-31G* basis set. The corresponding potential energies, gradients and non-Born-Oppenheimer couplings necessary for the classical evolution of the nuclear coordinates and solution of the nuclear Schrödinger equation in the time-evolving basis set of Gaussian wavepackets centered on classical trajectories are computed on the fly.[24-27] The coupled electron-nuclear dynamics of $C_2D_2^{2+}$ is simulated with the AIMS method[22] (see Supplementary Note 3B), which solves the electronic and nuclear Schrödinger equations simultaneously using a basis set of travelling Gaussian wavepackets for the nuclear wave function and determining the electronic structure as needed with the CASSCF method.[28,29] Although nonadiabatic crossing effects can be described by AIMS, we found that the short time (sub-100fs) dynamics of the dication after Auger decay is almost entirely adiabatic and the population of each state can be considered as constant on this time scale.

As shown in Supplementary Figure 6a, trans-bending of the dication is energetically highly disfavored on the low-lying $1\pi_u^{-2}$ states. In contrast, trans-bending up to ~60° is possible on the $1\pi_u^{-1}3\sigma_g^{-1}$ states. The different behavior of these electronic states can be understood from the nature of the bonding. The $\sigma$-bond is located along the C-C axis, and favors a linear C-C-D structure. Removing an electron from the bonding $\sigma$-orbital weakens the $\sigma$-bond, leading to more freedom for the deuterons to bend such that the $\theta_{CCD}$ angle is increased. Energetically, large amplitude bending (which could lead to isomerization) of the deuterons is possible on the $1\pi_u^{-1}3\sigma_g^{-1}$ states, and it is conceivable that this could take place on the sub-100fs timescale. However, in these states it is also energetically favorable to break the C-C bond (Supplementary Figure 6b) and dissociation could compete with isomerization. For the lowest three $1\pi_u^{-2}$ states, the large isomerization barrier (≈2eV) suggests that isomerization on these states is highly unlikely to complete on sub-100fs time scale. Thus, for all the low-lying singlet states, we might expect ultrafast isomerization to be a rare channel. On the $1\pi_u^{-2}$ states the barrier is too high and on the $1\pi_u^{-1}3\sigma_g^{-1}$ states, the symmetric breakup or deprotonation channels might be more likely. As shown in Figure 3, the simulations (corresponding to an ensemble of 500 trajectory basis functions for singlet and triplet states) clearly do not observe any isomerization in the first 100fs of the dication dynamics (recall the preparation of the dication follows from direct modeling of the X-ray pump and subsequent Auger decay). Figures 3a and 3b depict the time evolution of the probability distribution for the deuteron-coordination number $N_D(C_i)$ of the left and right carbon atoms, defined as:[30]

$$N_D(C_i) = \sum_j S\left(\left|r_{D_j} - r_{C_i}\right|\right), \quad i,j = a,b \qquad (1)$$

where

$$S(r) = 1/(\exp[\kappa(r - r_c)] + 1) \quad (2)$$

and we choose $r_c$=1.4Å and $\kappa^{-1}$=0.1Å. The coordination number $n_C$ provides a smoothed count of the number of deuterons within bonding distance of each carbon atom. In no case does this coordination number exceed one, which is a clear sign that no isomerization occurs on this timescale. Isomerization would entail one of the carbon atoms having a coordination number near two, while the other would have a coordination number near zero. For example, the CASSCF-optimized geometry of $CCD_2^{2+}$ on the $^1\Delta_g$ electronic state yields coordination numbers $N_D(C_{left})$ and $N_D(C_{right})$ of 0.0 and 1.9, respectively. Figure 3c shows the four C-D distances as a function of time, grouped such that the longer two such distances (at time t=0) are colored in red and the shorter two are colored in green. Again, it is quite clear that there is no switching of a deuteron between carbon atoms. This would appear to be inconsistent with the experiments that showed apparent evidence of ultrafast isomerization. In order to resolve this conundrum, we also simulated the Coulomb explosion process so that we could compare directly to the experimental observables. It is worth noting that the experimental analysis of Coulomb explosion data often assumes that the explosion is so fast that all rotation stops immediately. Under this approximation, the momentum map which would be expected from the experiment if there was no evolution on the dication state (neither isomerization nor dissociation) is shown in Figure 3d (details in Supplementary Note 3). Although the sudden approximation is clearly not strictly valid, as has been observed previously,[18] it is necessary to connect the measured Coulomb explosion momentum mapping data to a unique structure.

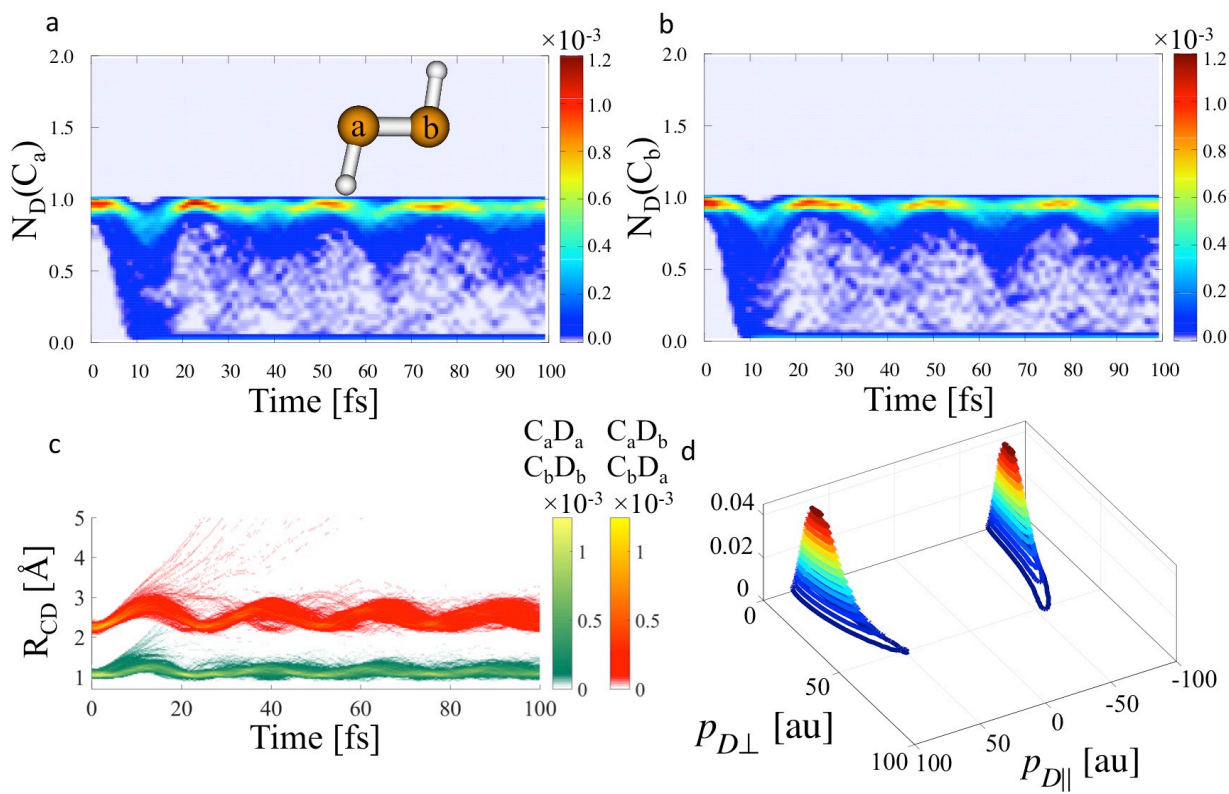

**Figure 3. Dynamics of acetylene dication after Auger decay.** Deuteron coordination number ($N_D$) of the two carbon atoms (a) $C_a$ and (b) $C_b$. See main text for definition of the deuteron coordination number. (c) The C-D distances ($R_{CD}$) for the initially bonded $C_a$-$D_a$ and $C_b$-$D_b$ atoms (green color), and for the initially nonbonded $C_a$-$D_b$ and $C_b$-$D_a$ atoms (orange color). (d) Effective momentum distribution produced from non-dissociating trajectories, assuming no remnant rotation between momentum vectors during the Coulomb explosion. $p_{D\perp}$ and $p_{D\parallel}$ are the components of deuteron momenta that are perpendicular and parallel to the C-C axis, respectively.

Following the recent experiment,[5] we simulated the Coulomb explosion momentum map (CEMM) image that was used to measure the nuclear motion. At each delay time and for each of the trajectory basis functions from the dication dynamics, we placed point charges (with appropriate masses) at the locations of each of the nuclei and propagated classically to simulate the experiment directly. Figure 4a shows the time evolution of the momentum map, which can be compared to Figure 3 of Ref. 5. Note that the region with large momentum perpendicular to the C-C axis and small momentum along the C-C axis fills in as the time delay increases. This was the primary observation in Ref. 5 leading to the conclusion that isomerization was occurring. In order to make the comparison more clear, we also plot the kinetic energy release distribution as a function of ∠CCD ($\theta_{CCD}$) in Figure 4c, again following the experimental data analysis. The natural assumption here is that population with ∠CCD less than $\pi/2$ are vinylidene-like, i.e. isomerized. Figure 4c can be compared directly with Figure 2 of Ref. 5 and (as in the experiment) shows considerable signal in the vinylidene like region (to the left of the black line in Figure 4c). This is in spite of the fact that there is no isomerization in the simulation data. Furthermore, the redshift in KER for the V-like channel observed in the experiment is also seen in the simulations (this redshift is due to kinetic energy loss when there is trans-bending of the deuteron). The V-like signal with low KER originates from higher energy trajectories with significant proton motion. Due to angular momentum conservation, rotation continues during Coulomb explosion for both the $C_2D^+ + D^+$ deprotonation and $CD^+ + CD^+$ symmetric breakup channels. The Coulomb explosion of the rotating fragments leads to V-like signals in the momentum mapping (Supplementary Figure 10 and Supplementary Note 3). Thus, while these channels do contain significant deuteron motion, they do not result in isomerization and would not break up into $C^+/CD_2^+$ had they not been Coulomb exploded. In Supplementary Figure 7, we show the dynamics of C-C axis rotation which was used as a clock in an earlier acetylene dication experiment that was also interpreted to support ultrafast isomerization.[12] Due to significant C-C bond elongation, its rotation decelerates. Supplementary Figure 7 shows that the C-C axis rotation depends linearly on time for only the first 30fs. After that time, the C-C axis rotation is nearly time-independent for the 100fs of our simulations. This implies that the clock is only accurate up to ≈30fs and the true upper bound for the isomerization reaction is not 60fs but rather some longer time scale (which is difficult to quantify without an improved model to calibrate the clock).

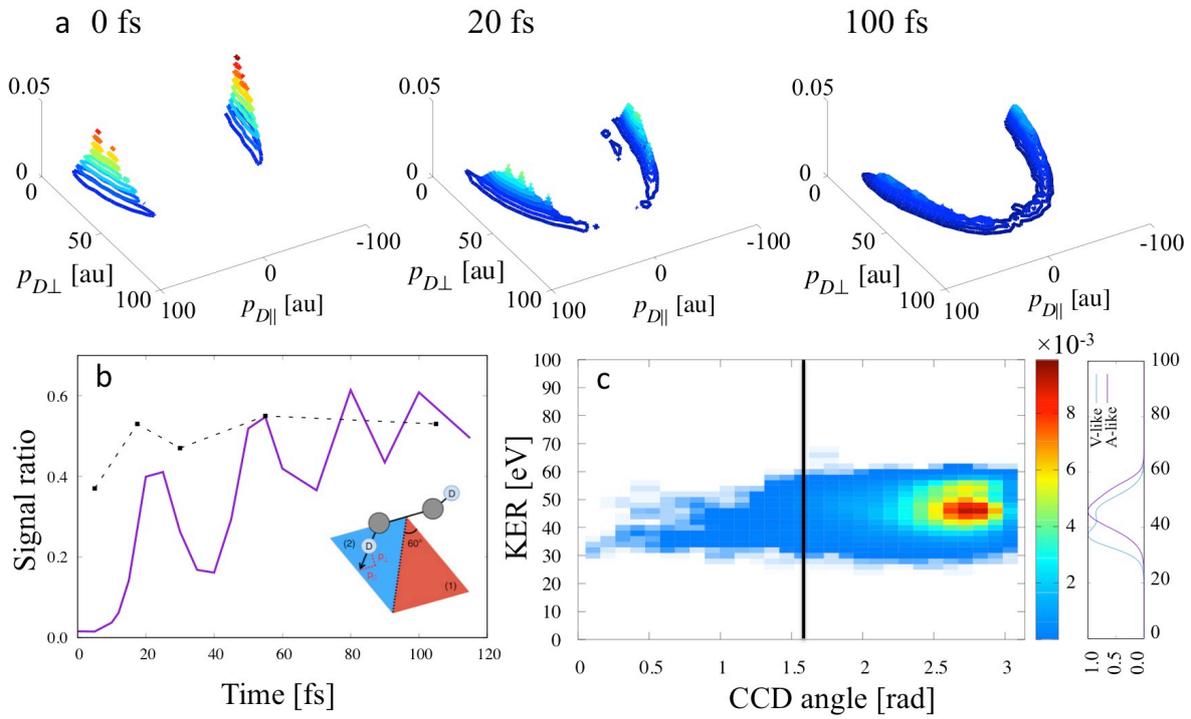

**Figure 4. Simulated Coulomb explosion momentum mapping signal.** (a) Temporal evolution of the deuteron momenta from 0 to 100 fs (axes correspond to directions parallel and perpendicular to the C-C axis) with an assumed instrumental broadening of $3.8 \times 10^{-22}$ *kg m s$^{-1}$* (b) Ratio of signal from red-shaded region where $0° \leq \tan^{-1}(p_\parallel / p_\perp) < 60°$ and blue-shaded region where $60° \leq \tan^{-1}(p_\parallel / p_\perp) < 120°$. Solid line is from simulations, which is compared with the experiments (dashed line). (c) The total kinetic energy release (KER) and angular distribution of C+/C+/D+/D+ coincidences integrated over all time delays. The CCD angle is defined as $\tilde{\theta} = \cos^{-1}\left( \dfrac{\text{sgn}\left[(p_{C_a} - p_{C_b}) \cdot p_{D_2}\right]\left((p_{C_a} - p_{C_b}) \cdot p_{D_1}\right)}{|p_{C_a} - p_{C_b}||p_{D_1}|} \right)$, as the angle a deuteron momentum makes with the effective C-C axis.[5]

Although the AIMS simulations do not support the isomerization channel, isomerization could take place with very low probability. Because the non-Born-Oppenheimer coupling allows transitions between electronic states induced by nuclear motion,[8,24,27,31-33] the dication can undergo partial proton migration (large $\theta_{CCD}$ angle) on the $1\pi_u^{-1}3\sigma_g^{-1}$ states which have a soft (nearly flat) angular potential, and then decay downwards via an electronic transition to the $1\pi_u^{-2}$ states where it might complete the isomerization. This nonadiabatically-assisted isomerization mechanism depends on the electronic transition occurring after the isomerization barrier on the lower $1\pi_u^{-2}$ states. Such a trajectory is shown in Supplementary Figure 6c with the trans-bending motion to $\theta_{CCD}$ angle of ~80° within 60 fs on the state $S_4$ and $S_3$ with $1\pi_u^{-1}3\sigma_g^{-1}$ or $1\pi_u^{-2}+\pi_u$-$\pi_g^*$ character, and it completes isomerization on $S_2$ at ~90 fs. However, nearly all trajectory basis functions with high trans-bending angles on states $S_3$ and $S_4$ do not follow this path but instead fragment along the C-C bond because of the high vibrational energy in the C-C stretch and the barrierless character of the potential along this coordinate (Supplementary Figure 6b). Using a simplified model (see Supplementary Note 3C), we can estimate the branching ratio of the nonadiabatically-assisted isomerization channel and the kinematically favored C-C symmetric fragmentation channel to be ~$1\times10^{-4}$. This estimate can be compared to the estimate of $1\times10^{-6}$ from the AIMS simulations.

We further point out that isomerization is possible from the satellite state $^1\Sigma_u$ ($S_7$) with 3-hole-1-particle (3h1p) electronic character $1\pi_u^{-2}+\pi_u$-$\pi_g^*$. The $^1\Sigma_u$ state is accessed from the shake-up state $(C_{1s})^{-1}+\pi_u$-$\pi_g^*$ in photoionization with Auger energy of 256.8 eV. It crosses the $^1\Pi_u$ and $^1\Pi_g$ states, and could switch to these states that support ultrafast isomerization. Using the sudden approximation, we can estimate the $K$-shell photoionization cross section as $\sigma \sim \left|\left\langle \Psi_{M^+}^{N-1}\left|\hat{a}\right|\Psi_{M^0}^{N}\right\rangle\right|^2$ and determine the ratio of the shake-up state $(C_{1s})^{-1}+\pi_u$-$\pi_g^*$ to the $(C_{1s})^{-1}$ state to be ~0.03. Due to their extremely low probability, these channels alone are insufficient to explain the abundant vinylidene-like signals found in the experiment.

To further investigate the possibility of an isomerization pathway, we take advantage of the fact that Coulomb explosion imaging is sensitive to the actual C-C bond length at the time of tetracation generation. An unbroken C-C bond is expected to result in V-like signal with higher KER due to larger Coulomb potential of short C-C distance, from which we can remove the symmetric breakup channel with significant hydrogen migration. We specifically look at the momentum difference of coincident carbon ions $p_{C^+C^+} = |p_{C_a^+} - p_{C_b^+}|$. We could not identify any signal in the experimental data set[5] beyond the simulated CEMM signals from non-isomerized trajectories at sufficient confidence level, as detailed in Supplementary Note 3. We can thus conclude with high confidence that isomerization is not occurring with any significant probability in acetylene dication prepared by Auger decay after X-ray core ionization.[5]

As a final note, we also observe the signature of vibrational coherence in the bending motion of the acetylene dication, as seen in Figure 4b. Since the vibrational motion of the acetylene dication is synchronized by the X-ray pump pulse when the vibrational frequency is suddenly changed (by ionization), vibrational coherence can be expected to occur for 100fs before dephasing. This is also known as a squeezed vibrational state,[5,34] analogous to the squeezed coherent state of photons. The vibrational coherence manifests itself in the ratio of deuteron with large bending angles (in region (1), Figure 4b) and in the vicinity of carbon atom (in region (2)), with a period of ~27 fs, which is half of the period of trans-bending motion and is consistent with a squeezed vibrational state that gives collective vibrational amplitude proportional to $\left[\left(1+(\omega_0/\omega_1)^2\right)+\left(1-(\omega_1/\omega_0)^2\right)\cos(2\omega_1 t)\right]^{1/2}$, where $\omega_0$ and $\omega_1$ are the vibrational frequencies before and after the pump pulse (details in Supplementary Note 3E).

Our study resolves the long-standing controversy between experiment and theory concerning the mechanism of the purported sub-100fs isomerization of acetylene dication. We conclude that in fact what appeared as ultrafast isomerization in previous experiments is actually significant proton migration on the ground state, or on excited states that then decays into symmetric breakup. Isomerization, which requires a stable C-C bond, can only occur in the low-lying states of the dication if the molecules have enough internal energy (and time) to overcome the isomerization barrier. This mechanism is infeasible on the sub-100fs time scale of the pump-probe experiment modeled here. Enough energy may be available for isomerization after non-adiabatic internal conversion from high-lying dicationic states towards the low energetic states. However, in this case, direct symmetric fragmentation dominates overwhelmingly.

Our work calls for cautious interpretation of the widely used Coulomb explosion momentum mapping (CEMM) method when resolving the transient geometry of molecular motion on femtosecond time scale. On the other hand, it also highlights CEMM's ability to resolve the ultrafast dynamics of momentum dispersion. Even when no significant geometric variation takes place, CEMM reveals the rich dynamics of the momentum distribution that changes substantially on the femtosecond time scale. With complementary transient geometry information from single molecule diffraction, which is enabled by X-ray free electron lasers or relativistic electrons,[35,36] we could form a complete picture of molecular dynamics in the entire phase space including both position and momentum. Such a time-resolved diffraction study was recently reported[37] and we expect that the combination of the simulations reported here, the previous CEMM measurements, and time-resolved diffraction will give a complete picture of the femtosecond dynamics of acetylene dication.

## DATA AVAILABILITY STATEMENT

The datasets generated during and/or analysed during the current study are available from the corresponding author on request.

## ACKNOWLEDGEMENT

This work was supported by the AMOS program within the Chemical Sciences, Geosciences and Biosciences Division of the Office of Basic Energy Sciences, Office of Science, US Department of Energy and the Hamburg Center for Ultrafast Imaging. Z. L. thanks the Volkswagen Foundation for support through a Peter Paul Ewald postdoctoral fellowship. Z. L. thanks Lee-Ping Wang, Koudai Toyota, Sang-Kil Son, Robin Santra, Daniel Haxton, Daniel Neumark, Mohamed El-Amine Madjet, Kota Hanasaki, Victor Kimberg and Yajiang Hao for stimulating discussions.

## AUTHOR CONTRIBUTIONS STATEMENT

T.M. and Z.L. conceived the concept and methods. T.M., Z.L., L.I. B.C., J.S., N.M., S.P. and O.V. conducted the theoretical investigation. C.L.-S., J.C., T.O. and P.B. conducted the experimental data analysis. All authors analyzed the results and reviewed the manuscript.

**COMPETING FINANCIAL INTERESTS STATEMENT**

The authors declare no competing financial interests.

# References

1   Levin, J., Feldman, H., Baer, A., Ben-Hamu, D., Heber, O., Zajfman, D. & Vager, Z. Study of Unimolecular Reactions by Coulomb Explosion Imaging: The Nondecaying Vinylidene. *Phys. Rev. Lett.* **81**, 3347-3350, (1998).

2   Jensen, F., Pedersen, U. V. & Andersen, L. H. Stability of the Ground State Vinylidene Anion H2CC−. *Phys. Rev. Lett.* **84**, 1128-1131, (2000).

3   Pople, J. A., Frisch, M. J., Raghavachari, K. & Schleyer, P. v. R. The structure and stability of the acetylene dication. *J. Comput. Chem.* **3**, 468-470, (1982).

4   Carrington, T., Hubbard, L. M., Schaefer, H. F. & Miller, W. H. Vinylidene: Potential energy surface and unimolecular reaction dynamics. *J. Chem. Phys.* **80**, 4347-4354, (1984).

5   Liekhus-Schmaltz, C. E., Tenney, I., Osipov, T., Sanchez-Gonzalez, A., Berrah, N., Boll, R., Bomme, C., Bostedt, C., Bozek, J. D., Carron, S., Coffee, R., Devin, J., Erk, B., Ferguson, K. R., Field, R. W., Foucar, L., Frasinski, L. J., Glownia, J. M., Guhr, M., Kamalov, A., Krzywinski, J., Li, H., Marangos, J. P., Martinez, T. J., McFarland, B. K., Miyabe, S., Murphy, B., Natan, A., Rolles, D., Rudenko, A., Siano, M., Simpson, E. R., Spector, L., Swiggers, M., Walke, D., Wang, S., Weber, T., Bucksbaum, P. H. & Petrovic, V. S. Ultrafast isomerization initiated by X-ray core ionization. *Nature Comm.* **6**, 8199, (2015).

6   Ibrahim, H., Wales, B., Beaulieu, S., Schmidt, B. E., Thiré, N., Fowe, E. P., Bisson, É., Hebeisen, C. T., Wanie, V., Giguére, M., Kieffer, J.-C., Spanner, M., Bandrauk, A. D., Sanderson, J., Schuurman, M. S. & Légaré, F. Tabletop imaging of structural evolutions in chemical reactions demonstrated for the acetylene cation. *Nature Comm.* **5**, 4422, (2014).

7   Jiang, Y. H., Rudenko, A., Herrwerth, O., Foucar, L., Kurka, M., Kühnel, K. U., Lezius, M., Kling, M. F., van Tilborg, J., Belkacem, A., Ueda, K., Düsterer, S., Treusch, R., Schröter, C. D., Moshammer, R. & Ullrich, J. Ultrafast Extreme Ultraviolet Induced Isomerization of Acetylene Cations. *Phys. Rev. Lett.* **105**, 263002, (2010).

8   Madjet, M. E.-A., Vendrell, O. & Santra, R. Ultrafast Dynamics of Photoionized Acetylene. *Phys. Rev. Lett.* **107**, 263002, (2011).

9   Jiang, Y. H., Senftleben, A., Kurka, M., Rudenko, A., Foucar, L., Herrwerth, O., Kling, M. F., Lezius, M., Tilborg, J. V. & Belkacem, A. Ultrafast dynamics in acetylene clocked in a femtosecond XUV stopwatch. *J. Phys. B At. Mol. Opt. Phys.* **46**, 164027, (2013).

10  Kübel, M., Siemering, R., Burger, C., Kling, N. G., Li, H., Alnaser, A. S., Bergues, B., Zherebtsov, S., Azzeer, A. M., Ben-Itzhak, I., Moshammer, R., de Vivie-Riedle, R. & Kling, M. F. Steering Proton Migration in Hydrocarbons Using Intense Few-Cycle Laser Fields. *Phys. Rev. Lett.* **116**, 193001, (2016).

11  Madjet, M. E.-A., Li, Z. & Vendrell, O. Ultrafast hydrogen migration in acetylene cation driven by non-adiabatic effects. *J. Chem. Phys.* **138**, 094311, (2013).

12  Osipov, T., Cocke, C. L., Prior, M. H., Landers, A., Weber, T., Jagutzki, O., Schmidt, L., Schmidt-Böcking, H. & Dörner, R. Photoelectron-Photoion Momentum Spectroscopy as a Clock for Chemical Rearrangements: Isomerization of the Di-Cation of Acetylene to the Vinylidene Configuration. *Phys. Rev. Lett.* **90**, 233002, (2003).

13  Osipov, T., Rescigno, T. N., Weber, T., Miyabe, S., Jahnke, T., Alnaser, A. S., Hertlein, M. P., Jagutzki, O., Schmidt, L. P. H., Schöffler, M., Foucar, L., Schössler, S., Havermeier, T., Odenweller, M., Voss, S., Feinberg, B., Landers, A. L., Prior, M. H., Dörner, R., Cocke, C. L. & Belkacem, A. Fragmentation pathways for selected electronic states of the acetylene dication. *J. Phys. B* **41**, 091001, (2008).

14  Zyubina, T. S., Dyakov, Y. A., Lin, S. H., Bandrauk, A. D. & Mebel, A. M. Theoretical study of isomerization and dissociation of acetylene dication in the ground and excited electronic states. *J. Chem. Phys.* **123**, 134320, (2005).

15  Duflot, D., Robbe, J.-M. & Flament, J.-P. Ab initio study of the acetylene and vinylidene dications fragmentation. *J. Chem. Phys.* **102**, 355-363, (1995).

16  Hishikawa, A., Matsuda, A., Fushitani, M. & Takahashi, E. J. Visualizing recurrently migrating hydrogen in acetylene dication by intense ultrashort laser pulses. *Phys. Rev. Lett.* **99**, 258302, (2007).

17  Flammini, R., Fainelli, E., Maracci, F. & Avaldi, L. Vinyldene dissociation following the Auger-electron decay of inner-shell ionized acetylene. *Phys. Rev. A* **77**, 044701, (2008).

18  Matsuda, A., Fushitani, M., Takahashi, E. J. & Hishikawa, A. Visualizing hydrogen atoms migrating in acetylene dication by time resolved three body and four body Coulomb explosion imaging. *Phys. Chem. Chem. Phys.* **13**, 8697-8704, (2011).

19  B. Gaire, S. Y. Lee, D. J. Haxton, P. M. Pelz, I. Bocharova, F. P. Sturm, N. Gehrken, M. Honig, M. Pitzer, D. Metz, H.-K. Kim, M. Schöffler, R. Dörner, H. Gassert, S. Zeller, J. Voigtsberger, W. Cao, M. Z., J. Williams, A. Gatton, D. Reedy, C. Nook, Thomas Müller, A. L. Landers, C. L. Cocke, I. Ben-Itzhak, T. Jahnke, A. Belkacem & Weber, T. Photo-double-ionization of ethylene and acetylene near threshold. *Phys. Rev. A* **89**, 013403, (2014).

20  Gilbert, A. T. B., Besley, N. A. & Gill, P. M. W. Self-Consistent Field Calculations of Excited States Using the Maximum Overlap Method. *J. Phys. Chem. A* **112**, 13164-13171, (2008).

21  Siegbahn, H., Asplund, L. & Kelfve, P. The Auger electron spectrum of water vapour. *Chem. Phys. Lett.* **35**, 330-335, (1975).

22  Ben-Nun, M. & Martínez, T. J. Ab Initio Quantum Molecular Dynamics. *Adv. Chem. Phys.* **121**, 439-512, (2002).

23  Roos, B. O. The Complete Active Space Self-Consistent Field Method and its Applications in Electronic Structure Calculations. *Adv. Chem. Phys.* **69**, 399-445, (1987).

24  Martinez, T. J., Ben-Nun, M. & Ashkenazi, G. Classical/quantal method for multistate dynamics: A computational study. *J. Chem. Phys.* **104**, 2847-2856, (1996).

25  Martinez, T. J., Ben-Nun, M. & Levine, R. D. Multi-Electronic-State Molecular Dynamics: A Wave Function Approach with Applications. *J. Phys. Chem.* **100**, 7884-7895, (1996).

26  Martinez, T. J. & Levine, R. D. First-principles molecular dynamics on multiple electronic states: A case study of NaI. *J. Chem. Phys.* **105**, 6334-6341, (1996).

27  Bearpark, M. J., Bernardi, F., Clifford, S., Olivucci, M., Robb, M. A., Smith, B. R. & Vreven, T. The Azulene S1 State Decays via a Conical Intersection:  A CASSCF Study with MMVB Dynamics. *J. Amer. Chem. Soc.* **118**, 169-175, (1996).

28  Snyder, J. W., Hohenstein, E. G., Luehr, N. & Martinez, T. J. An atomic orbital-based formulation of analytical gradients and nonadiabatic coupling vector elements for the state-averaged complete active-space self-consistent field method on graphical processing units. *J. Chem. Phys.* **143**, 154107, (2015).

29    Hohenstein, E. G., Luehr, N., I. S. Ufimtsev & Martinez, T. J. An atomic orbital-based formulation of the complete active space self-consistent field method on graphical processing units. *J. Chem. Phys.* **142**, 224103, (2015).

30    Sprik, M. Coordination Numbers as Reaction Coordinates in Constrained Molecular Dynamics. *Faraday Disc.* **110**, 437-445, (1998).

31    Barbatti, M., Shepard, R. & Lischka, H. in *Conical Intersections: Theory, Computation and Experiment*   (eds W. Domcke, D. R. Yarkony, & H. Köppel)   415-462 (World Scientific, 2011).

32    Domcke, W. & Yarkony, D. R. Role of Conical Intersections in Molecular Spectroscopy and Photoinduced Chemical Dynamics. *Ann. Rev. Phys. Chem.* **63**, 325-352, (2012).

33    Worth, G. A. & Cederbaum, L. S. Beyond Born-Oppenhemer: Molecular Dynamics Through a Conical Intersection. *Ann. Rev. Phys. Chem.* **55**, 127-158, (2004).

34    Johnson, S. L., Beaud, P., Vorobeva, E., Milne, C. J., Murray, É. D., Fahy, S. & Ingold, G. Directly Observing Squeezed Phonon States with Femtosecond X-Ray Diffraction. *Phys. Rev. Lett.* **102**, 175503, (2009).

35    Glownia, J. M., Natan, A., Cryan, J. P., Hartsock, R., Kozina, M., Minitti, M. P., Nelson, S., Robinson, J., Sato, T., van Driel, T., Welch, G., Weninger, C., Zhu, D. & Bucksbaum, P. H. Self-Referenced Coherent Diffraction X-Ray Movie of Angstrom- and Femtosecond-Scale Atomic Motion. *Phys. Rev. Lett.* **117**, 153003, (2016).

36    Yang, J., Guehr, M., Shen, X., Li, R., Vecchione, T., Coffee, R., Corbett, J., Fry, A., Hartmann, N., Hast, C., Hegazy, K., Jobe, K., Makasyuk, I., Robinson, J., Robinson, M. S., Vetter, S., Weathersby, S., Yoneda, C., Wang, X. & Centurion, M. Diffractive Imaging of Coherent Nuclear Motion in Isolated Molecules. *Phys. Rev. Lett.* **117**, 153002, (2016).

37    Wolter, B., Pullen, M. G., Le, A. T., Baudisch, M., Doblhoff-Dier, K., Senftleben, A., Hemmer, M., Schroter, C. D., Ullrich, J., Pfeifer, T., Moshammer, R., Grafe, S., Vendrell, O., Lin, C. D. & Biegert, J. Ultrafast electron diffraction imaging of bond-breaking in di-ionized acetylene. *Science* **354**, 308-312, (2016).

**Supplementary Note 1. The electronic structure and dynamics of the core-ionized acetylene cation**

We model the electronic structure of the core ionized acetylene cation [DCCD]$^+$ with a ΔSCF method[1] using the PBE0 exchange-correlation functional and the 6-31G* basis set. The core-hole is assumed to be localized on one of the carbon atoms due to coupling with non-totally symmetric modes.[2] The electronic potential energies and gradients are computed using GAMESS. In short, a closed-shell Kohn-Sham equation is solved to provide the initial orbitals for ΔSCF orbital optimization. The core shell orbitals are then localized on each carbon atom, respectively. A self-consistent field (SCF) procedure then follows with a singly occupied C1s core orbital. The SCF procedure is carried out with the maximum overlap method to preserve the character of the core hole singly-occupied orbital. In the maximum overlap method, the occupied orbitals in each new SCF cycle are reordered and phased to maximize the overlap with the occupied orbitals of previous SCF cycle. The overlap matrix of MOs from the $k$-th and $k+1$-th SCF cycles is defined as

$$O_{ij}^{k,k+1} = \left\langle \varphi_i^k \middle| \varphi_j^{k+1} \right\rangle = \sum_{pq} C_{ip}^k S_{pq} C_{jq}^{k+1} \qquad (1)$$

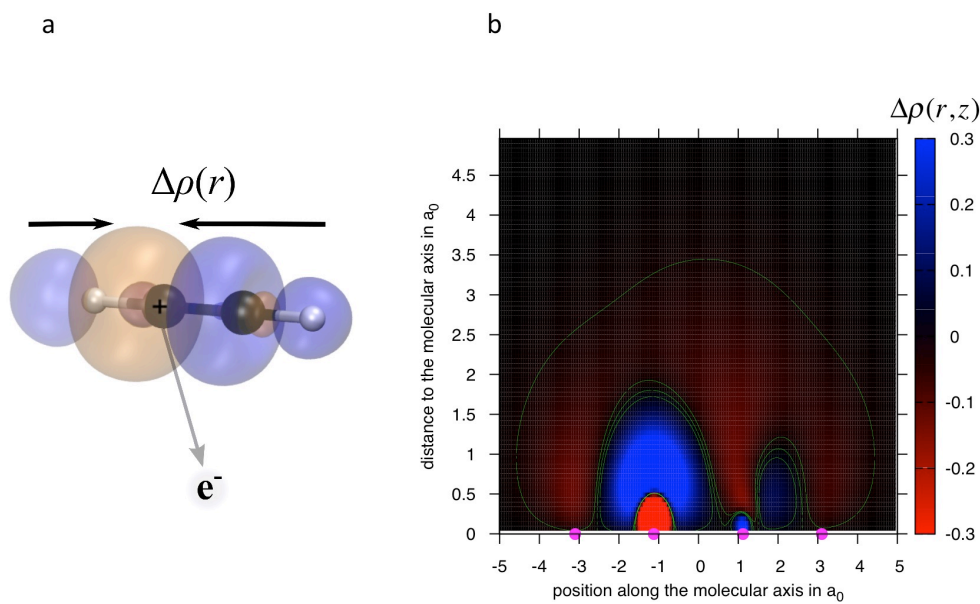

**Supplementary Figure 1. Electronic screening after core ionization.** (a) Sketch of the screening dynamics. The C1s core hole, as a positive charge, attracts valence electrons in the electronic relaxation after core photoionization. The orange/blue surfaces represent positive/negative increments in the total electron density $\Delta\rho(\mathbf{r})$ between the fully relaxed electronic structure, $\rho_{screen}$, and that with a pure C1s core hole in Koopmans' sense, $\rho_0$. (b) 2D illustration of $\Delta\rho(r,z)$ in cylindrical coordinates, the positions of carbon and deuterium atoms are labeled with dots on the x-axis, and $\Delta\rho(r,z) = 2\pi r\left(\rho_{screen}(r,z) - \rho_0(r,z)\right)$.

The ordering and phase of the MOs is adjusted to maximize the coincidence of the MOs between the $k$-th and $k+1$-th SCF cycles. The phase can be adjusted for any of the MOs (multiplying by ±1), but reordering is only allowed for orbitals with the same occupation number.

The dynamic screening effect of valence electrons is shown in Supplementary Figure 1, depicting the difference of the electron density between the core-ionized cation (with one electron removed from a localized C1s orbital) and the neutral molecule. We can estimate the time scale of the corresponding shake-up core hole relaxation after core-ionization and build-up of dynamical correlation from the C1s electron binding energy using the time-energy uncertainty principle as $t = 2\pi / E_{1s} \approx 25$ as. As shown in Supplementary Figure 1a, the net effect of core hole relaxation is to increase the valence bonding electron density in the region of the C-C bond, an increase which is due to attraction by the positively charged core hole. The C-C bond is thus strengthened, and the acetylene cation [HCCH]$^+$ has a shorter equilibrium C-C bond length than the neutral acetylene molecule, as shown in Supplementary Figure 2e. It is also a direct consequence of the valence electron screening effect that the angular potential along $\theta_{CCD}$ coordinate is hardened at shorter C-C bond length $R_{CC}$, because bonding along the C-C axis is strengthened (see Supplementary Figure 2b). However, the angular potential for fixed C-C bond length shows a slight softening (see Supplementary Figures 2c and 2d).

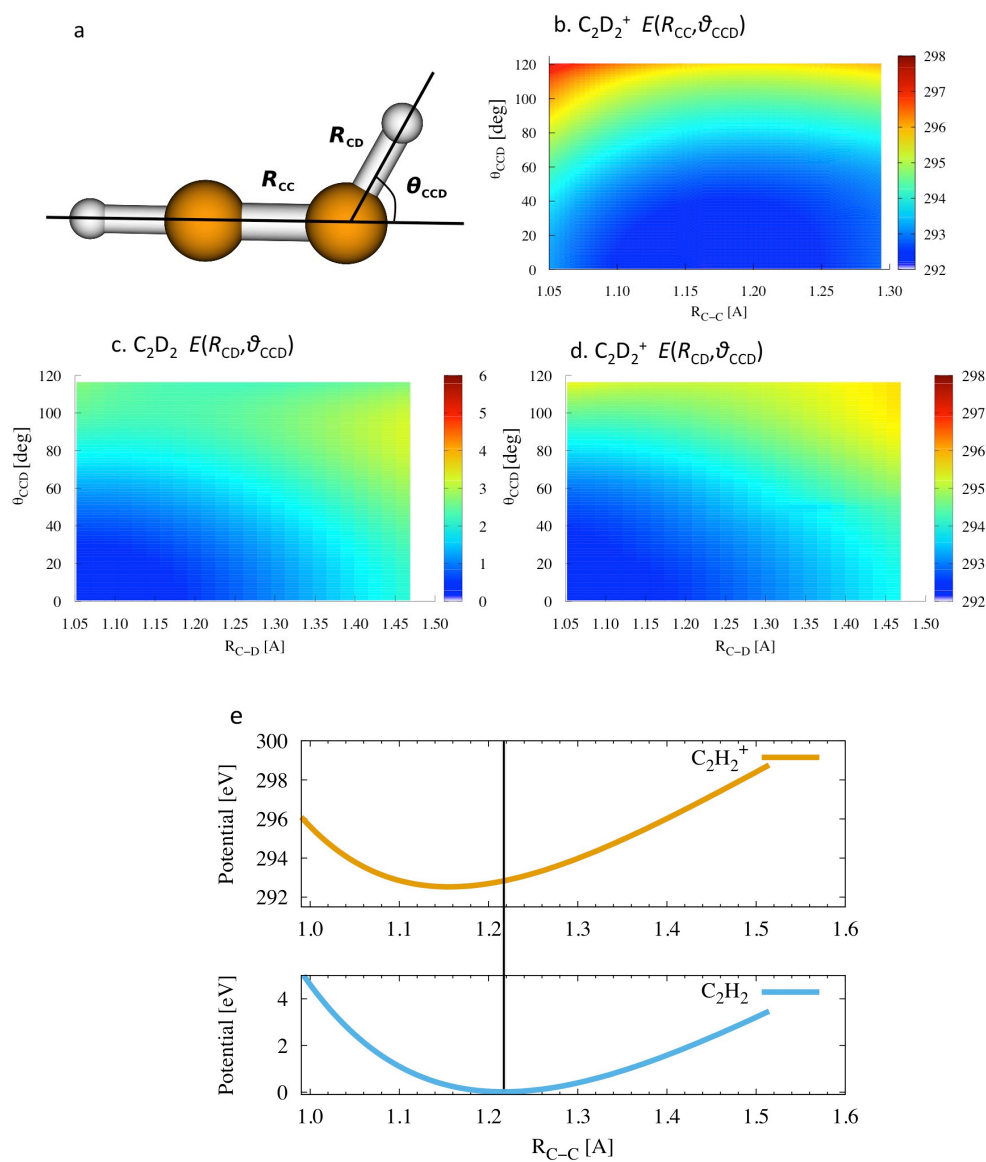

**Supplementary Figure 2. Potential energy surface of the $^2\Sigma$ C1s core ionized acetylene cation.** (a) Illustration of the coordinates $R_{CC}$, $R_{CD}$ and $\theta_{CCD}$. (b) 2D potential $E(R_{CC}, \theta_{CCD})$ of core ionized acetylene cation with $R_{CD}$=1.06Å. (c) 2D potential $E(R_{CD}, \theta_{CCD})$ of neutral acetylene molecule on $^1\Sigma_g$ state with $R_{CC}$=1.22Å. (d) 2D potential $E(R_{CD}, \theta_{CCD})$ of core ionized acetylene cation on $^2\Sigma$ state with $R_{CC}$=1.22Å. (e) 1D potential $E(R_{CC})$ of core ionized acetylene cation on $^2\Sigma$ state with $R_{CD}$=1.06Å and $D_{\infty h}$ symmetry.

**Supplementary Note 2. Auger decay and generation of acetylene dication**

The Auger decay rates are computed along the trajectories from a single Slater determinant wave function built with Kohn-Sham orbitals. The amplitude of Auger decay with an initial core hole in orbital $\phi_c$ to a final electronic configuration with double valence hole in orbitals $\phi_a$, $\phi_b$ and an Auger electron wave function $\phi_k$ with momentum $k$ can be determined as[3]

$$A(\mathrm{i}[c] \to \mathrm{f}[ab][k]) \sim \left( \langle ab|ck \rangle \delta_{\sigma_a \sigma_c} \delta_{\sigma_b \sigma_k} - \langle ba|ck \rangle \delta_{\sigma_b \sigma_c} \delta_{\sigma_a \sigma_k} \right) \langle \Psi_f^N | \hat{c}_a^\dagger \hat{c}_b^\dagger \hat{c}_c | \Psi_i^{N-1} \rangle |k\rangle, \quad (2)$$

where $\langle ab|ck \rangle$ is defined as

$$\langle ab|ck \rangle = \iint dr_1 dr_2 \, \phi_a(r_1)\phi_b(r_2) \frac{1}{|r_1 - r_2|} \phi_c(r_1)\phi_k(r_2). \quad (3)$$

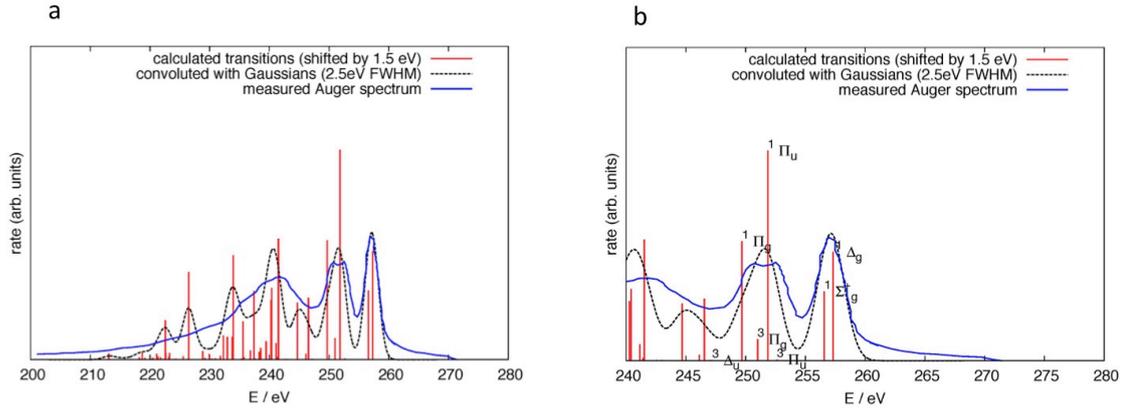

**Supplementary Figure 3. KLL Auger spectrum of acetylene molecule.** (a) Spectrum over wide energy range, simulated using single-center expansion method at equilibrium geometry subject to primary carbon K-shell photoionization.[4] The broadening width is taken to be 2.5 eV in order to match the experimental[5] data. The spectrum is dominated by the Auger decay into singlet dicationic states, with an Auger lifetime of ~8 fs. (b) Same spectrum as in (a), but zoomed in to a narrower energy range and with assignments for lowest energy peaks.

Based on the fact that the core hole orbital $\phi_c$ is strongly localized on the atom subject to primary photoionization, the molecular continuum wave function $\phi_k$ in the two-electron integral can be approximated with the atomic continuum wave function $\phi_{\varepsilon lm}$ with $\varepsilon$ the energy of the atomic Auger transition and $l,m$ the angular quantum numbers. Employing this approximation, the two electron integrals are approximated using the LCAO expansion only on the atom on which the core hole is localized[6]

$$\langle ab|ck\rangle = \sum_{\mu\nu\lambda,\,\text{on atom}} \sum_{lm} C_{\mu a} C_{\nu b} C_{\lambda c} \langle \mu\nu|\lambda;\varepsilon lm\rangle, \quad \varepsilon = \varepsilon_\mu + \varepsilon_\nu - \varepsilon_\lambda = \frac{k^2}{2}. \quad (4)$$

Assuming frozen molecular orbitals and neglecting any configurational mixing, the Auger rates are calculated for the singlet (S) and triplet (T) final states as

$$\Gamma_S\left(\text{i}[c] \to \text{f}[ab][\varepsilon]\right) = 2\pi \sum_{lm} N_{ab} \left[ \frac{1}{2} \left| \langle ab|c;\varepsilon lm\rangle + \langle ba|c;\varepsilon lm\rangle \right|^2 \right],$$

$$\Gamma_T\left(\text{i}[c] \to \text{f}[ab][\varepsilon]\right) = 2\pi \sum_{lm} N_{ab} \left[ \frac{3}{2} \left| \langle ab|c;\varepsilon lm\rangle - \langle ba|c;\varepsilon lm\rangle \right|^2 \right], \quad (5)$$

where $N_{ab} = 1, a \neq b$ and $N_{ab} = 1/2, a = b$. For the evaluation of atomic two electron integrals we employ the XATOM code.[7] The use of a minimal basis allows a direct identification of basis functions with atomic orbitals.

A simulated Auger spectrum at the Franck-Condon geometry is presented in Supplementary Figure 3. The spectrum is computed with the single center expansion method,[4] where the double hole states after Auger decay are described by wave functions determined from multi-reference configuration interaction (MRCI).

From the simulated Auger spectrum in Supplementary Figure 3, we could determine the FWHM of the experimental spectra to be ~2.5eV[5]. In the coincidence experiment,[8] the Auger electrons that correlate with the putative isomerization channel have a distribution centered at 255.5 eV.[8] These Auger electrons could have three sources, from the low energy wing of the $1\pi_u^{-2}$ states ($^1\Delta_g$, $^1\Sigma_g$) centered at 257.1±0.2eV, from the high energy wing of the $1\pi_u^{-1}3\sigma_g^{-1}$ states ($^1\Pi_u$) centered at 252.6±0.4eV, and from the 3h1p satellite state ($^1\Sigma_u$) with Auger energy of 256.8 eV that is accessed via photoionization and shake-up process. The pathways on the $^1\Pi_u$ and $^1\Sigma_u$ states were assumed to support the purported ultrafast isomerization on the sub-100 fs time scale,[9] and the pathways on the $^1\Delta_g$ and $^1\Sigma_g$ states are open for isomerization on a longer time scale that corresponds to the dynamics of potential barrier crossing. These channels all together contribute to the measured vinylidene-like signals in Refs. 8,10.

For dynamical Auger spectra, we adopt the single determinant ΔSCF method for computational efficiency. The Auger spectra show dominant decay channels into singlet states, and a cumulative Auger lifetime of ~8 fs. Within the Auger lifetime, it is sufficient for the geometrical relaxation to take place on the cationic $^2\Sigma$ state, that the acetylene cation evolves towards a linear structure with shorter C-C bond length. The Auger rates computed at each step along the trajectories are used to construct initial conditions of dynamics for the dication through solution of the corresponding kinetic rate equation.

**Supplementary Note 3. Dynamics on the dicationic states and Coulomb explosion imaging**

A. Dicationic potentials

The electronic structure of acetylene dication during the ab initio multiple spawning molecular dynamics is computed using the dynamically weighted[11] SA8-CASSCF(8,8)/6-31G* method, which is quantitatively consistent with MS8-CASPT2(8,10)/6-31G*. In Supplementary Figure 4, we present the diabatic potential of the dication along the trans-bending coordinate $\theta_{CCD}$. The potential curve of the $^1\Sigma_u$ state cuts through those of the $^1\Pi_g$ and $^1\Pi_u$ states as the C-C bond stretches (Figure 1 in the main text). The $^1\Sigma_u$, $^1\Pi_g$ and $^1\Pi_u$ states together form a low-barrier pathway for isomerization in the corresponding adiabatic states.

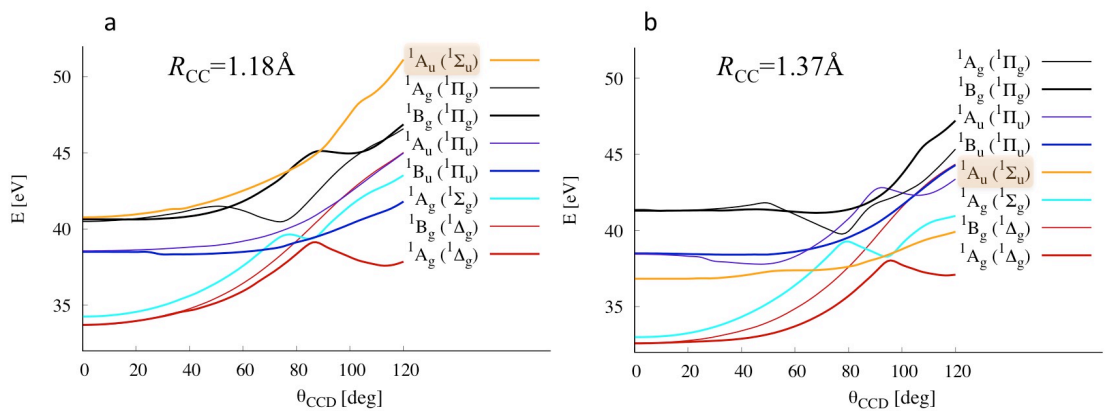

**Supplementary Figure 4. 1D diabatic potentials $E(\theta_{CCD})$ of acetylene dication.** Potential energy surfaces are calculated at the SA8-CASSCF(8,8)/6-31G* level, with $R_{CD}$=1.12Å and $R_{CC}$=1.18Å, 1.37Å, i.e. the C-C bond lengths of the Franck-Condon geometry and the equilibrium C-C bond length of the dicationic ground state. The electronic characters are labeled by the representations of $C_{2h}$ symmetry and that correlated with linear geometry (in parenthesis, $\theta_{CCD}$ =0°, $D_{\infty h}$ symmetry). At the Franck-Condon geometry, the $^1\Delta_g$ and $^1\Sigma_g$ states that correlate with adiabatic states $S_0$-$S_2$ are of $1\pi_u^{-2}$ character, the $^1\Pi_u$ and $^1\Pi_g$ states that correlate with adiabatic states $S_3$-$S_4$ are of electronic character $1\pi_u^{-1}3\sigma_g^{-1}$ and $1\pi_u^{-1}2\sigma_u^{-1}$. The $^1\Sigma_u$ state is a 3-hole-1-particle (3h1p) satellite state with electronic character $1\pi_u^{-2}+(\pi_u \rightarrow \pi_g^*)$, it correlates initially with $S_7$ at Franck-Condon geometry (a), and becomes $S_3$ as the C-C bond stretches (b). It is noticeable that the potential of $^1\Sigma_u$ state flattens along the $\theta_{CCD}$ coordinate as the C-C bond relaxes. The zero of energy is set as the ground state energy of acetylene at its equilibrium geometry.

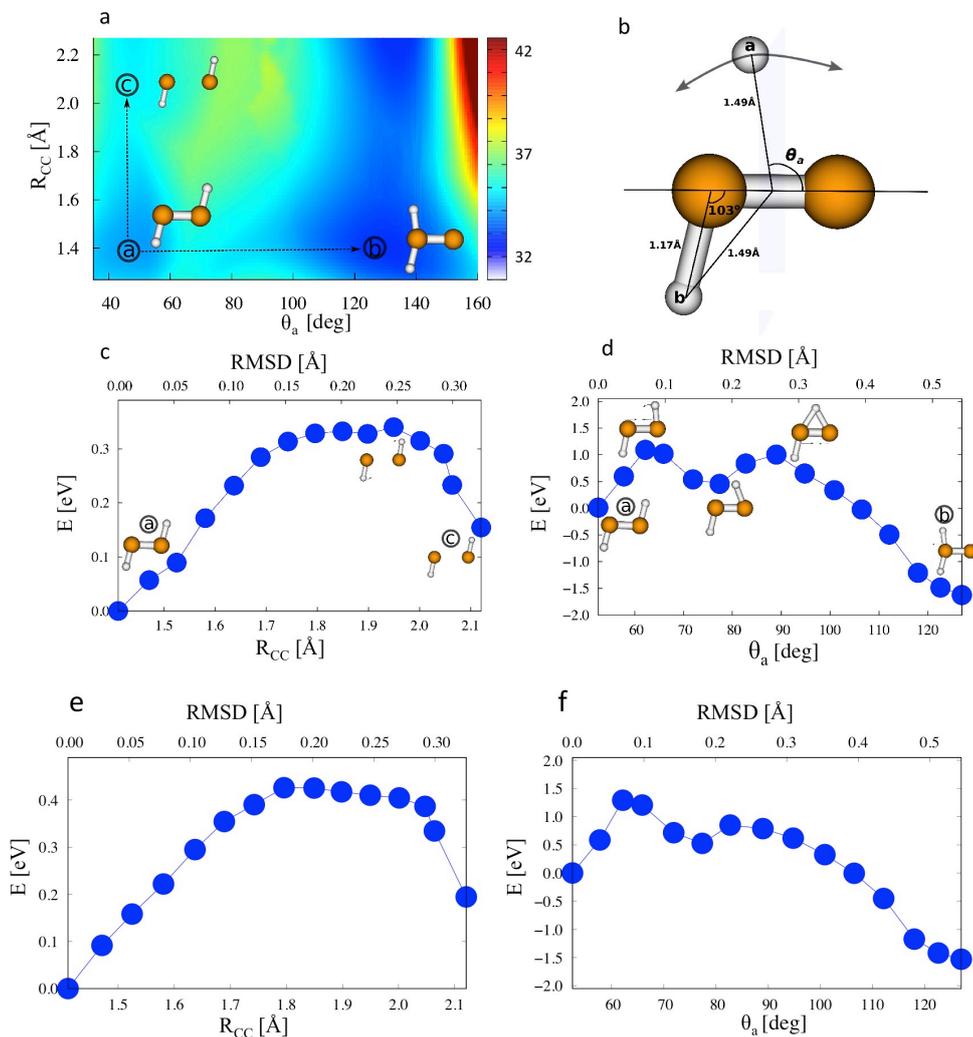

**Supplementary Figure 5. Isomerization pathway on the $1\pi_u^{-2}$ state $S_2$.** Geometry changes on $S_2$ are crucial to complete the isomerization when the dication decays from $1\pi_u^{-1}3\sigma_g^{-1}$ states. (a) 2D potential $E(R_{CC},\theta)$ from scan with rigid geometry on the remaining degrees of freedom. (b) Illustration of degrees of freedom for rigid potential scan in (a), deuteron $D_b$ is fixed at the position from optimized geometry of $[CD_2C]^{2+}$. (c) Potential along C-C fragmentation pathway with relaxed geometry. (d) Potential along isomerization pathway with relaxed geometry. (e), (f) Potential along C-C fragmentation (e) and isomerization (f) pathways with relaxed geometry at the MS8-CASPT2(8,10) level.

In Supplementary Figure 5, we present the potential along C-C fragmentation and isomerization pathways on the $1\pi_u^{-2}$ state $S_2$, which is crucial to complete isomerization process by providing a bounded C-C potential for the dication. We optimize the geometry on the excited dicationic states for the isomerized and fragmented dication, and calculate the potential along the isomerization and fragmentation coordinates. Supplementary Figure 5c shows a bound potential for the C-C stretch coordinate with a barrier of ~0.3 eV, and in Supplementary Figure 5d, we show that isomerization requires formation of a triangular C-D-C bridge as the deuteron migrates halfway to the middle of the two carbon atoms, with the two carbon atoms contracting towards each other. This however imposes another disadvantageous factor to the isomerization channel, because the C-C stretch motion is kinematically enhanced due to kinetic energy release from geometric relaxation after Auger decay. Benchmark calculations at the MS8-CASPT2(8,10)/6-31G* level in Supplementary Figures 5e and 5f show the reliability of the *ab initio* potentials obtained on the SA8-CASSCF(8,8)/6-31G* level for the AIMS dynamics.

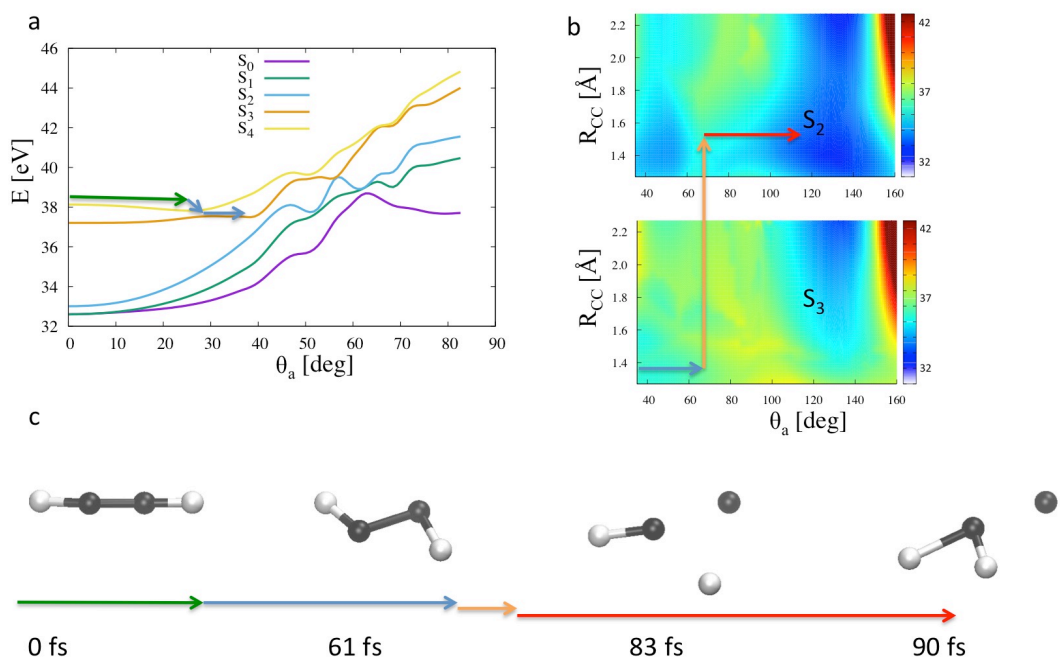

**Supplementary Figure 6. Schematic of a possible sub-100 fs isomerization pathway.** (a) trans-bending potential plotted with $\theta_a$ coordinate and assuming $\theta_b = \pi + \theta_a$, where the angle $\theta_a$ is described in Supplementary Figure 5b. The trajectory (c) starts from the high lying $1\pi_u^{-1}3\sigma_g^{-1}$ states $S_4$ (a) that allows the deuteron to move to a large angle before 60 fs, the non-Born-Oppenheimer transition downwards to $S_3$ and $S_2$ offers a bounded potential for the C-C bond (b), the dication then isomerizes on the low lying $1\pi_u^{-2}$ states at ca. 90 fs. In (a) the potentials are calculated with $R_{CD} = 1.12$Å, $R_{CC} = 1.37$Å and trans-bending angle $\theta_{CCD}$, in (b) they are calculated with $R_{CM-D} = 1.49$Å, $R_{CC} = 1.37$Å, and deuteron migration angle $\theta_a$, $R_{CM-D}$ is the distance of the migrating hydrogen to the center of mass (CM) of the two carbon atoms, and we define $\theta_a = \theta_{C-CM-D}$. The arrow in (a) illustrates the direction of the isomerization, and the arrow in (b) shows the pathway of electronic decay through the conical intersection from $S_3$ to $S_2$ that occur at the end of the isomerization process. Note that although this pathway is possible, the simulations predict it to be highly improbable.

B. Ab initio multiple-spawning (AIMS) molecular dynamics simulation

The coupled electron-nuclear dynamics of acetylene dication is simulated with the ab initio multiple spawning (AIMS) method,[12] which solves the electronic and nuclear Schrödinger equations simultaneously using a travelling Gaussian wavepacket basis for the nuclear wave function and electronic structure methods at various levels (dynamically-weighted SA-CASSCF in this work). The total electronic nuclear wave function is represented by a time-dependent basis set of atom-centered Gaussians for the nuclear degrees of freedom as

$$\Psi(R,r,t) = \sum_I \sum_i^{N_I(t)} c_i^I(t) \chi_i^I(R; \bar{R}_i^I, \bar{P}_i^I, \gamma_i^I) \phi_I(r;R), \qquad (6)$$

where I labels electronic states, $N_I(t)$ is the number of nuclear basis functions associated with the Ith electronic state at time t, and r/R label electronic/nuclear coordinates respectively. The nuclear basis increases during the simulation when nonadiabatic coupling increases and electronic transitions are possible, which is dubbed *spawning*, and the nuclear Schrödinger equation is solved in this basis to determine the time evolution of electronic population. The nuclear basis functions $\chi_i^I(R; \bar{R}_i^I, \bar{P}_i^I, \gamma_i^I)$ are parameterized by the positions/momenta in phase space and a phase factor $\gamma_i^I$. The electronic basis functions $\phi_I(r;R)$ are solved using the GPU-based electronic structure program TeraChem[13,14] in the adiabatic representation at the SA8-CASSCF(8,8)/6-31G* level to provide electronic potential energies, gradients and non-Born-Oppenheimer couplings at each step. The complex coefficients $c_i^I(t)$ evolve following the time-dependent Schrödinger equation (TDSE) in the time-dependent basis, i.e.:

$$\sum_{kK} S_{jk}^{JK} \dot{c}_k^K = -i \sum_{kK} (H_{jk}^{JK} - i\dot{S}_{jk}^{JK}) c_k^K \tag{7}$$

The overlap matrix, its time derivative and the Hamiltonian matrix elements are defined as:

$$\begin{aligned} S_{jk}^{JK} &= \langle \chi_j^J \phi_J | \chi_k^K \phi_K \rangle \delta_{JK}, \\ \dot{S}_{jk}^{JK} &= \left\langle \chi_j^J \phi_J \left| \frac{\partial \chi_k^K}{\partial t} \phi_K \right. \right\rangle \delta_{JK}, \\ H_{jk}^{JK} &= \langle \chi_j^J \phi_J | \hat{H} | \chi_k^K \phi_K \rangle. \end{aligned} \tag{8}$$

The observables O(t), such as positions and momenta of atoms, for the Ith electronic state are calculated as

$$\langle O(t) \rangle_I = \frac{\sum_k^{N_I(t)} c_k^{I*} c_k^I \langle \chi_k^I \phi_k | \hat{O} | \chi_k^I \phi_k \rangle}{\sum_k^{N_I(t)} c_k^{I*} S_{kk}^{II} c_k^I} \tag{9}$$

The observables averaged over all electronic states can be obtained from incoherent summation (for operators which are diagonal in the electronic state index) over the electronic states as

$$\langle O(t) \rangle = \sum_I \rho_I(t) \langle O(t) \rangle_I \tag{10}$$

where $\rho_I(t)$ is the population of *I*th electronic state at time *t*.

C. Dynamics of dication

The recoil momentum of the outgoing Auger electron with energy of ca. 255 eV and the angular momentum of the trans-bending motion can set the $[C_2D_2]^{2+}$ dication into rotation, and this rotation has been used as a clock to estimate the time of isomerization.[10]

In Supplementary Figure 7, we show the rotational motion of the C-C axis from the molecular dynamics simulation. The existence of rotation is consistent with the previous observation, however the dications only rotate for ~0.15 rad in 100 fs, slower than expected in Ref. 10, and the dispersion of the $\theta_{CC}$ angle distribution also shows a deceleration. Because the C-C bond stretches from 1.17Å to 1.37Å, the angular velocity ω decelerates due to conservation of angular momentum. Additionally, the recoil momentum from an Auger electron of 255 eV adds at most ~0.012 eV of kinetic energy to the bending motion of the dication when it acts normal to the C-C axis. The recoil motion has very limited influence on isomerization in comparison with the total kinetic energy of 0.95 eV, and it is especially hard to compete with the 0.43 eV of kinetic energy in the C-C stretching motion (which favors the symmetric fragmentation channel). For the lowest three singlet states where the σ-orbital remains doubly occupied, the unidirectional σ-bond along the C-C axis should weaken the ability of recoil momentum to cause isomerization. Furthermore, the Auger electron from delocalized valence orbitals will impose recoil momentum to all atoms, further diminishing its ability to promote isomerization. Thus we suggest that the isomerization due to barrier crossing on the lower $1\pi_u^{-2}$ dicationic state takes place at much longer time scales, which is consistent with the conclusion from an energetic perspective,[15] and also with the observed dynamics in AIMS simulations.

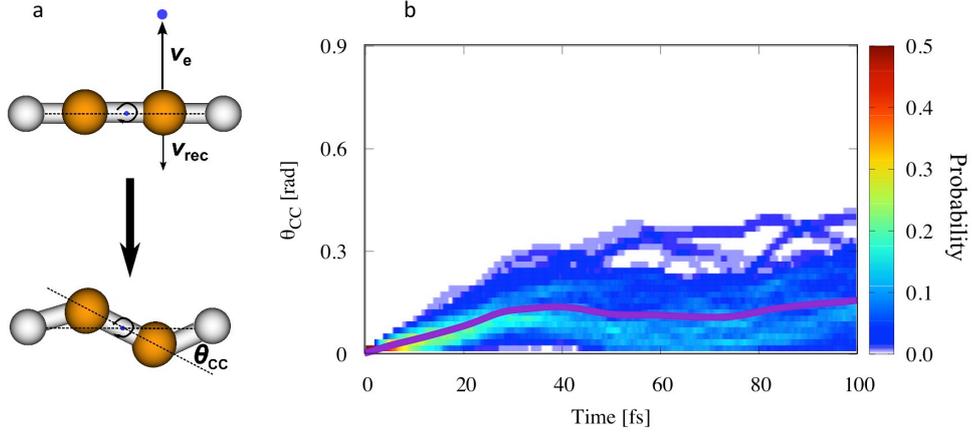

**Supplementary Figure 7. Rotational motion of C-C axis after Auger decay.** (a) Sketch of rotation of C-C axis with initial velocity $v_{rec}$ due to recoil momentum of Auger electron with velocity $v_e$ on the lever, and possible supporting effect the C-C rotational motion to isomerization. (b) Rotation of C-C axis in the lab system due to recoil momentum of Auger electron, calculated from molecular dynamics simulation, where the rotation angle is defined as

$$\theta_{CC}(t) = \left| \cos^{-1}\left( \frac{\vec{R}_{CC}(t) \cdot \vec{R}_{CC}(0)}{|R_{CC}(t)||R_{CC}(0)|} \right) \right|.$$ The solid curve corresponds to $\langle \theta_{CC} \rangle (t)$.

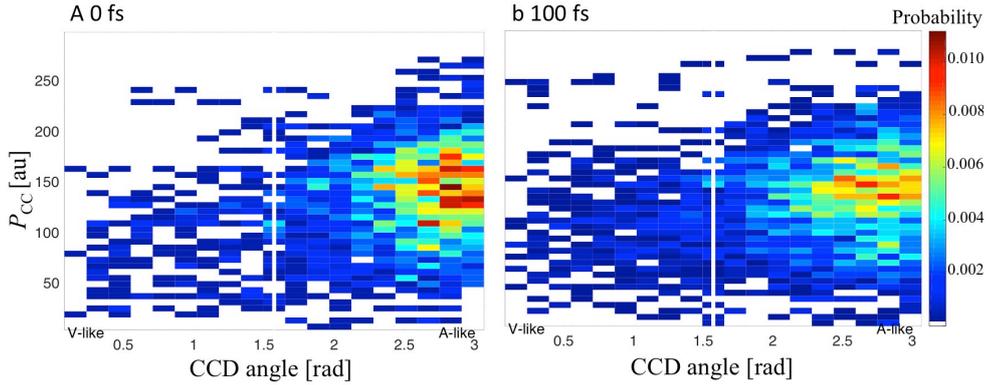

**Supplementary Figure 8. Four-particle coincidence Coulomb momentum mapping.** Experimental momentum difference data for $C^+$ ions in four-particle coincidence Coulomb momentum mapping. (a) coincidence counts at 0fs. (b) coincidence counts at 100fs.

In Figure 3 of the main text, we present geometric configurations of the dication that can be accessed by the ensemble of nuclear trajectories from the ab initio multiple spawning molecular dynamics simulation, where no significant isomerization can be detected. The most compelling argument against isomerization is that direct simulation does not produce significant isomerization while the resulting data is simultaneously in agreement with the experimental observations. However, it is also of interest to show that this result can be rationalized with a simplified analysis, which we do here. We calculated the kinetic energy distribution in the C-C stretching separately and the remaining 5 degrees of freedom, because the C-C stretching mode is the dominantly excited mode in the dication formation. Because both C-C stretch and trans-bending motion are quasi barrier free on the $1\pi_u^{-1}3\sigma_g^{-1}$ states, we can assume all the occupied states of these two degrees of freedom correspond to open channels. We can estimate the branching ratio between C-C fragmentation and isomerization simply from probability of open channels $P_i = g\frac{(2\pi m_i kT)^{3/2}V}{h^3}e^{-\frac{E_i}{kT}}$, where $m_i$ and $g$ are the reduced mass and degeneracy of the corresponding mode, respectively, and $V$ is the quantization volume. Assuming ambient temperature $T$=300 K, the kinetic energy of C-C stretch is obtained from the AIMS trajectories for the 100 fs dynamics as $\langle E_{KER,CC}\rangle = \frac{1}{T}\int dt\, E_{KER,CC}(t) \sim 0.43\,\text{eV}$, while the total kinetic energy is given as $\langle E_{KER}\rangle = \frac{1}{T}\int dt\, E_{KER}(t) \sim 0.95\,\text{eV}$. The reduced masses for C-C stretch and trans-bending motion are taken as 7 amu and 2.592 amu from spectroscopic data.[16] For simplicity, we assume all recoil energy of 0.012 eV is pumped into the trans-bending degrees of freedom (DOF), which will overestimate the likelihood of trans-bending. The rest of the kinetic energy $\langle E_{KER}\rangle - \langle E_{KER,CC}\rangle$ is equally distributed among the five DOFs. Of course, this is only an estimate since IVR is expected to complete on the picosecond timescale. We have then $E_{KER,TB} \sim \frac{2}{5}\times(0.52-0.012)+0.012=0.215\,\text{eV}$ for the trans-bending mode. The ratio between the isomerization and C-C fragmentation channels is then estimated to be $1.1\times10^{-4}$.

Figure 3d presents an angular distribution of the deuterons, which is plotted so that the eye can more easily compare it to the momentum distribution. First, the angular distribution is drawn on an annulus that is comparable the momentum distribution seen in the experiment. Then, the distribution is given a Gaussian width, again to mimic the momentum distribution seen in the experiment.

In principle, intersystem crossing might be relevant, and the isomerization might occur on a triplet state. In order to address this possibility, we have determined the largest spin-orbit coupling matrix element for acetylene dication at the SA8-CASSCF(8,8)/6-31G* level to be 3.21 meV between the $\left|{}^1\Delta_g; \Lambda = \pm1, M_S = 0, M_J = \pm1\right\rangle$ and $\left|{}^3\Pi; \Lambda = \pm1, M_S = 0, M_J = \pm1\right\rangle$ states, which corresponds to a time scale of 1.29ps. This makes intersystem crossing from the singlet to triplet states in acetylene dication dynamics quite improbable on the sub-100fs time scale. Here $\Lambda$, $M_S$ and $M_J$ are the projections of the orbital, spin and total angular momenta, respectively.

A further possibility to consider is the impact of tunneling on these results. We are confident that tunneling isomerization can be safely excluded from our scenario. Subject to a potential barrier of ~2eV and heavy deuteron isotope, the tunneling isomerization should take place on microsecond (μs) time scale,[17] which is substantially longer than the sub-100fs time scale.

D. Coulomb momentum mapping

The Coulomb momentum mapping by further core photoionization and Auger decay with an X-ray probe pulse of certain time delay produces a tetracation $[C_2D_2]^{4+}$. The Coulomb explosion of the tetracation is simulated by a classical Hamiltonian of four singly charged particles $C^+/C^+/D^+/D^+$,

$$H = T + V_{\text{Coul}}(r_1, r_2, r_a, r_b)$$
$$V_{\text{Coul}}(r_1, r_2, r_a, r_b) = \frac{1}{r_{1a}} + \frac{1}{r_{1b}} + \frac{1}{r_{ab}} + \frac{1}{r_{12}} + \frac{1}{r_{2a}} + \frac{1}{r_{2b}}, \quad (11)$$

because dynamics of the highly repulsive tetracation is not sensitive to its detailed electronic structure. In Eq. 12, $T$ is the kinetic energy, $r_1$, $r_2$, $r_a$, $r_b$ are the coordinates of the carbon and deuteron ions as shown in Supplementary Figure 9a, and $V_{\text{coul}}(r_1, r_2, r_a, r_b)$ is the Coulomb potential. An interatomic distance cutoff of 5 Å is set for atoms involved in the Coulomb explosion in order to guarantee the atoms subject to the same parent dication are charged by Auger decay induced by the X-ray probe pulse. Molecular dynamics of the tetracation $[C_2D_2]^{4+}$ is carried out for 5 ps. Final distances of ions are of ~$10^3$ Å, from which we can assume the Coulomb potential has been fully released. The deuteron momenta are projected on the C-C axis defined by the difference of momenta of the two carbon ions.

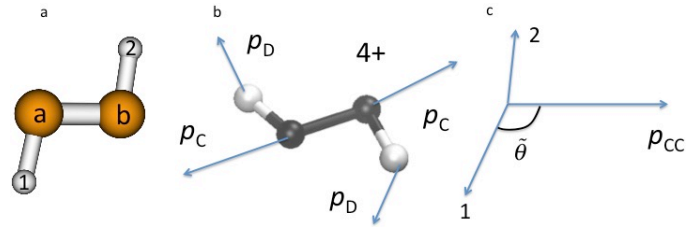

**Supplementary Figure 9. Schematic diagrams for Coulomb momentum mapping.** (a) Labeling of atoms in acetylene. (b) Labeling of momentum vectors. (c) Definition of angle which was used to separate acetylene-like and vinylidene-like signals (see Eq. 12).

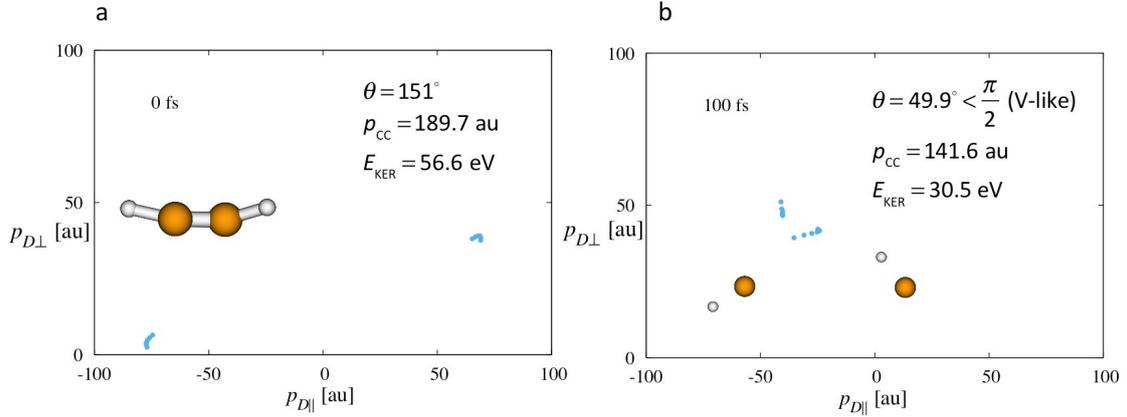

**Supplementary Figure 10. Simulated Coulomb momentum mapping.** Illustrative Coulomb momentum mapping data from a representative trajectory. X-ray probe pulse time delay of (a) t=0 fs and (b) t=100 fs. The trajectory gives V-like signal at time delay t=100 fs, even though it clearly belongs to the C-C fragmentation channel. The CD$^+$ ion fragment carries remnant angular momentum inherited from trans-bending motion before fragmentation. For each time delay t, we compute five deuteron momentum mappings from t-2 to t+2 fs.

To classify the four-particle coincidences signal, we define the angle $\tilde{\theta}$:[9]

$$\tilde{\theta} = \cos^{-1}\left(\frac{\operatorname{sgn}\left[\left(p_{C_a} - p_{C_b}\right) \cdot p_{D_2}\right]\left(\left(p_{C_a} - p_{C_b}\right) \cdot p_{D_1}\right)}{\left|p_{C_a} - p_{C_b}\right|\left|p_{D_1}\right|}\right) \tag{12}$$

We classify the C$^+$/C$^+$/D$^+$/D$^+$ four-particle-coincidence as vinylidene-like (V-like) for $\tilde{\theta} < \frac{\pi}{2}$, such that both deuteron momenta reside on the same side of the bisection plane that divides the C-C axis. The remaining coinicidences with $\tilde{\theta} > \frac{\pi}{2}$ are referred as acetylene-like (A-like) with deuteron momenta on opposite sides of the bisection plane. An important finding from our work is that V-like coincidence does not necessarily correspond to an isomerization event (at variance with previous assumptions).

Assuming a C-C bond length limit of dissociation to be 1.5 Å, and considering only the collinear Coulomb force between the two C$^+$ ions, an isomerized acetylene dication should give $p_{C^+C^+}$ with a lower bound of ~181 au and V-like signal in the four-particle momentum map. Because the $C^+$ ions also feel the Coulomb force from the $D^+$ ions, the actual $p_{C^+C^+}$ should be higher than 181 au. In the non-isomerized trajectories of the molecular dynamics calculation that satisfy the lower bound, the lowest $\tilde{\theta}$ angle they can reach is 20°, with $p_{C^+C^+} = 199$ au. In the V-like signals experimental data set, we attribute the four-particle coincidence signals with $\tilde{\theta} \leq 20°$ and $p_{C^+C^+} \geq 199$ au to the isomerization channel, which is not covered by the MD trajectories.

In Supplementary Figure 8, we show the momentum difference $p_{C^+C^+}$ of $C^+$ ions measured in the time-resolved x-ray pump x-ray probe experiment at LCLS.[9] The ab initio multiple spawning molecular dynamics simulation sets a criterion of for V-like signals that do not come from isomerization channel and reinforces the fact that pathways of sub-100 fs isomerization mediated by the non-Born-Oppenheimer effect is a very rare channel. In fact, only one initial condition (out of 500) ever approached this channel (with ∠CCD>75°), and the transition probability was $\approx 1\times 10^{-4}$. Thus, we estimate the probability of observing this channel to be less than $1\times 10^{-6}$. Indeed, the only reason to mention this channel is that it seems the only feasible way that any sub-picosecond isomerization could occur in the experiment.

In Supplementary Figure 10, we show a representative trajectory of symmetric fragmentation channel with breakup of C-C bond, which gives V-like coincidence signal due to the remnant angular momentum of the $CD^+$ fragments. However, the C-C fragmented trajectory with V-like coincidence should have lower KER and $p_{CC}$, and can be distinguished from isomerization channel with V-like coincidence, because long C-C distance simply implies low Coulomb potential release. Thus, we reconsider the measured Carbon momenta in the experiment with respect to the C-C momentum.

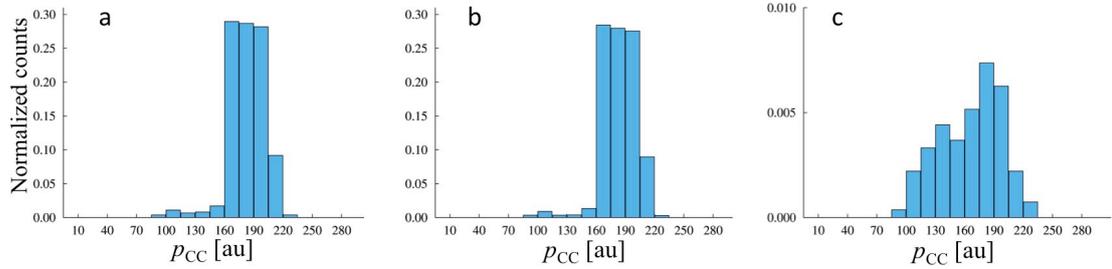

**Supplementary Figure 11. Simulated Coulomb momentum mapping.** Momentum difference of $C^+$ ions in the Coulomb momentum mapping at 100 fs from molecular dynamics calculations. (a) Total coinicidence counts. (b) Coincidence counts of A-like signals. (c) Coincidence counts of V-like signals.

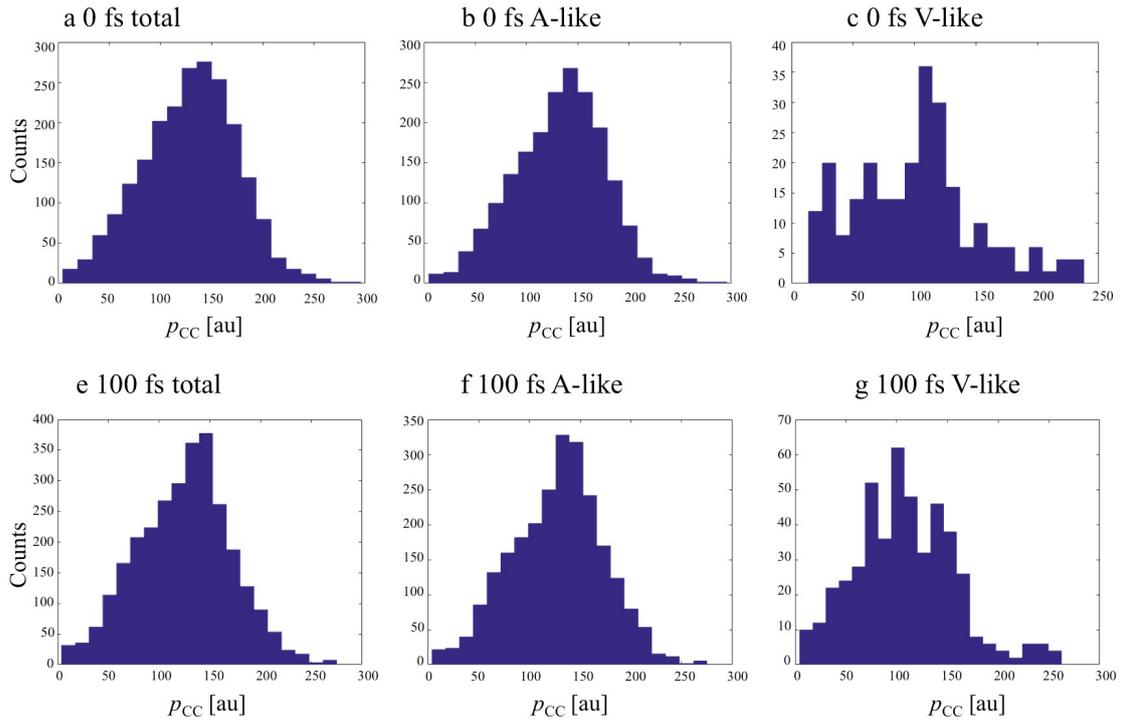

**Supplementary Figure 12. Experimental Coulomb momentum mapping.** Momentum difference of $C^+$ ions measured in Coulomb momentum mapping. Upper panels correspond to 0fs time delay and show (a) Total coinicidence counts, (b) coincidence counts of A-like signals, and (c) coincidence counts of V-like signals. Lower panels correspond to 100 fs time delay and show (e) Total coinicidence counts, (f) coincidence counts of A-like signals, and (g) coincidence counts of V-like signals.

Supplementary Figures 11 and 12 show the momentum difference obtained in the experiment and from the MD simulations.

It is noticeable that there are coincidence counts in Supplementary Figure 12 with low momentum difference $p_{C^+C^+} \leq 70$ au that appear within 100 fs, these coincidence counts are not found in the molecular dynamics simulation of 120 fs, and they correspond to C-C distance larger than 30 Å, these coincidence count signals may come from the highly dissociative states above $S_7$, which we have not considered in the simulation and should also have negligible contribution to isomerization. Especially, the coincidence counts with very low momentum difference down to 10 au appear from 0 fs (Supplementary Figure 11): these signals correspond to C-C distances that could be larger than 1 μm, and are comparable to the size of the XFEL-sample interaction region of 50 μm$^2$.$^9$ They give a clear hint that despite filtering using the momentum conservation criterion:

$$\left| p_{D_a,j} + p_{D_b,j} + p_{C_a,j} + p_{C_b,j} \right| = p_j < \delta p_j, j = x, y, z \tag{13}$$

in the lab system,$^9$ some of the coincidence counts of Coulomb momentum mapping may arise from $C^+D^+$ fragments of two distant $C_2D_2^{2+}$ dications instead of the same parent cation. These accidental coincidences must be further filtered out from the dataset in order to have a clean interpretation of the Coulomb momentum mapping experiment for the making of a molecular movie. To improve interpretation of the Coulomb momentum mapping technique, we need to introduce an additional criterion to ensure that the coincidence counts originate from fragments of the same parent ion by filtering out the coincidence with relative momentum of two ion particles which are lower than would be allowed by reasonable bond length between these two particles in their parent ion. This novel criterion basically reflects the essence of Coulomb explosion imaging of bond length.

E. Vibrational coherence produced by sudden change of frequency

After ionization, the vibrational frequencies will change due to the new electronic potential. This is the origin of the oscillation observed in Figure 4b of the main text (with a 27fs period, which is half the vibrational period of symmetric bending), as discussed here. Considering the coherent states that are closely correlated with classical particles in harmonic potential, the vibrational motion can be described by a time-dependent Hamiltonian

$$H = \hbar\omega(t)(a^+ a + \frac{1}{2}) ,  \qquad (14)$$

where

$$\omega(t) = \begin{cases} \omega_0, \ t \leq 0 \\ \omega_1, \ t > 0 \end{cases} , \qquad (15)$$

i.e. the frequency undergoes a sudden change at $t=0$ due to ionization. In the Heisenberg picture, we have

$$a(t) = e^{iHt/\hbar} a e^{-iHt/\hbar} = e^{-i\omega_1 t}\left( \frac{\omega_1+\omega_0}{2\sqrt{\omega_1\omega_0}} a + \frac{\omega_1-\omega_0}{2\sqrt{\omega_1\omega_0}} a^+ \right) = U(t)a + V(t)a^+ .$$
$$a^+(t) = U^*(t)a^+ + V^*(t)a$$

And we thus have for the collective vibrational amplitude

$$\begin{aligned}\langle Q^2 \rangle(t) &= \frac{\hbar}{2M\omega_1}\langle (a(t)+a^+(t))^2 \rangle \\ &= \frac{\hbar}{M\omega_1}\langle |U(t)+V^*(t)|^2 (n+\frac{1}{2}) \rangle \\ &= \frac{\hbar}{4M\omega_0}\coth\frac{\hbar\omega_0}{2k_B T}\left[ \left(1+(\omega_0/\omega_1)^2\right) + \left(1-(\omega_1/\omega_0)^2\right)\cos(2\omega_1 t) \right]\end{aligned} \qquad (16)$$

where $T$ is temperature, $M$ is reduced mass of the vibrational mode, and $k_B$ is the Boltzmann constant.

# Supplementary References

1. Gilbert, A. T. B., Besley, N. A. & Gill, P. M. W. Self-Consistent Field Calculations of Excited States Using the Maximum Overlap Method. *J. Phys. Chem. A* **112**, 13164-13171, (2008).
2. Gadea, F. X., Köppel, H., Schirmer, J., Cederbaum, L. S., Randall, K. J., Bradshaw, A. M., Ma, Y., Sette, F. & Chen, C. T. *Phys. Rev. Lett.* **66**, 883, (1991).
3. Manne, R. & Ågren, H. Auger transition amplitudes from general many-electron wavefunctions. *Chem. Phys.* **93**, 201-208, (1985).
4. Inhester, L., Burmeister, C. F., Groenhof, G. & Grubmüller, H. Auger spectrum of a water molecule after single and double core ionization. *J. Chem. Phys.* **136**, 144304, (2012).
5. Kivimäki, A., Neeb, M., Kempgens, B., Köppe, H. M., Maier, K. & Bradshaw, A. M. Angle-resolved Auger spectra of the C2H2 molecule. *J. Phys. B* **30**, 4279, (1997).
6. Siegbahn, H., Asplund, L. & Kelfve, P. The Auger electron spectrum of water vapour. *Chem. Phys. Lett.* **35**, 330, (1975).
7. Son, S.-K., Young, L. & Santra, R. Impact of hollow-atom formation on coherent x-ray scattering at high intensity. *Phys. Rev. A* **83**, 033402, (2011).
8. Osipov, T., Rescigno, T. N., Weber, T., Miyabe, S., Jahnke, T., Alnaser, A. S., Hertlein, M. P., Jagutzki, O., Schmidt, L. P. H., Schöffler, M., Foucar, L., Schössler, S., Havermeier, T., Odenweller, M., Voss, S., Feinberg, B., Landers, A. L., Prior, M. H., Dörner, R., Cocke, C. L. & Belkacem, A. Fragmentation pathways for selected electronic states of the acetylene dication. *J. Phys. B* **41**, 091001, (2008).
9. Liekhus-Schmaltz, C. E., Tenney, I., Osipov, T., Sanchez-Gonzalez, A., Berrah, N., Boll, R., Bomme, C., Bostedt, C., Bozek, J. D., Carron, S., Coffee, R., Devin, J., Erk, B., Ferguson, K. R., Field, R. W., Foucar, L., Frasinski, L. J., Glownia, J. M., Guhr, M., Kamalov, A., Krzywinski, J., Li, H., Marangos, J. P., Martinez, T. J., McFarland, B. K., Miyabe, S., Murphy, B., Natan, A., Rolles, D., Rudenko, A., Siano, M., Simpson, E. R., Spector, L., Swiggers, M., Walke, D., Wang, S., Weber, T., Bucksbaum, P. H. & Petrovic, V. S. Ultrafast isomerization initiated by X-ray core ionization. *Nature Comm.* **6**, 8199, (2015).
10. Osipov, T., Cocke, C. L., Prior, M. H., Landers, A., Weber, T., Jagutzki, O., Schmidt, L., Schmidt-Böcking, H. & Dörner, R. Photoelectron-Photoion Momentum Spectroscopy as a Clock for Chemical Rearrangements: Isomerization of the Di-Cation of Acetylene to the Vinylidene Configuration. *Phys. Rev. Lett.* **90**, 233002, (2003).
11. Glover, W. J. *J. Chem. Phys.* **141**, 171102, (2014).
12. Ben-Nun, M. & Martínez, T. J. Ab Initio Quantum Molecular Dynamics. *Adv. Chem. Phys.* **121**, 439-512, (2002).
13. Hohenstein, E. G., Luehr, N., I. S. Ufimtsev & Martinez, T. J. An atomic orbital-based formulation of the complete active space self-consistent field method on graphical processing units. *J. Chem. Phys.* **142**, 224103, (2015).
14. Snyder, J. W., Hohenstein, E. G., Luehr, N. & Martinez, T. J. An atomic orbital-based formulation of analytical gradients and nonadiabatic coupling vector elements for the state-averaged complete active-space self-consistent field method on graphical processing units. *J. Chem. Phys.* **143**, 154107, (2015).
15. Zyubina, T. S., Dyakov, Y. A., Lin, S. H., Bandrauk, A. D. & Mebel, A. M. Theoretical study of isomerization and dissociation of acetylene dication in the ground and excited electronic states. *J. Chem. Phys.* **123**, 134320, (2005).
16. Palaudoux, J., Jutier, L. & Hochlaf, M. Theoretical spectroscopy of acetylene dication and its deuterated species. *J. Chem. Phys.* **132**, 194301, (2010).
17. Scharma, A. R., Bowman, J. M. & Nesbitt, D. J. *J. Chem. Phys.* **136**, 034305, (2012).